\documentclass[longauth]{aa} 

\usepackage{graphicx}
\usepackage{txfonts}
\begin{document}

   \title{The luminous red nova variety: AT~2020hat and AT~2020kog}

   \subtitle{Forbidden hugs in pandemic times - II}

   \author{A.~Pastorello\inst{1}
          \and
          G.~Valerin\inst{1,2}
          \and
          M.~Fraser\inst{3}
          \and
          N.~Elias-Rosa\inst{1,4}
          \and
          S.~Valenti\inst{5}           
          \and
          A.~Reguitti\inst{6,7,1}
           \and 
          P.~A.~Mazzali\inst{8,9}
           \and
          R.~C.~Amaro\inst{10}
          \and
          J.~E.~Andrews\inst{10}
          \and
          Y.~Dong\inst{5}
           \and
          J.~Jencson\inst{10}
           \and
          M.~Lundquist\inst{10}
           \and
          D.~E.~Reichart\inst{11}
           \and
          D.~J.~Sand\inst{10}
          \and
          S.~Wyatt\inst{10}
           \and  
          S.~J.~Smartt\inst{12}
           \and
          K.~W.~Smith\inst{12}
         \and
          S.~Srivastav\inst{12}
            \and                           
          Y.-Z.~Cai\inst{13}
           \and    
          E. Cappellaro\inst{1}
           \and    
          S.~Holmbo\inst{14}  
           \and 
          A.~Fiore\inst{1,2}
          \and
          D.~Jones\inst{15,16}
          \and
          E.~Kankare\inst{17}
           \and
          E.~Karamehmetoglu\inst{14}
           \and
          P. Lundqvist\inst{18}
           \and 
          A.~Morales-Garoffolo\inst{19}
             \and       
          T.~M.~Reynolds\inst{17}
           \and     
          M.~D.~Stritzinger\inst{14}
          \and 
          S.~C.~Williams\inst{17,20}
            \and 
          K.~C.~Chambers\inst{21}
            \and                    
          T.~J.~L.~de~Boer\inst{21}
            \and 
          M.~E.~Huber\inst{21}
            \and 
          A.~Rest\inst{22,23}   
           \and 
          R.~Wainscoat\inst{21}
}

   \institute{INAF - Osservatorio Astronomico di Padova, Vicolo dell'Osservatorio 5, I-35122 Padova, Italy  \email{andrea.pastorello@inaf.it}
          \and
             Universit\'a degli Studi di Padova, Dipartimento di Fisica e Astronomia, Vicolo dell’Osservatorio 2, 35122 Padova, Italy
          \and
             School of Physics, O’Brien Centre for Science North, University College Dublin, Belfield, Dublin 4, Ireland         
          \and
             Institute of Space Sciences (ICE, CSIC), Campus UAB, Carrer de Can Magrans s/n, E-08193 Barcelona, Spain
           \and
             Department of Physics and Astronomy, University of California, 1 Shields Avenue, Davis, CA 95616-5270, USA 
          \and
             Departamento de Ciencias Fisicas, Universidad Andres Bello, Fernandez Concha 700, Las Condes, Santiago, Chile
          \and
             Millennium Institute of Astrophysics (MAS), Nuncio Monsenor S\'otero Sanz 100, Providencia, Santiago, Chile
         \and
             Astrophysics Research Institute, Liverpool John Moores University, ic2, 146 Brownlow Hill, Liverpool L3 5RF, UK
          \and
             Max-Planck Institut f\"ur Astrophysik, Karl-Schwarzschild-Str. 1, D-85741 Garching, Germany
          \and
             Steward Observatory, University of Arizona, 933 North Cherry Avenue, Rm. N204, Tucson, AZ 85721-0065, USA
          \and
             Department of Physics and Astronomy, University of North Carolina at Chapel Hill, Chapel Hill, NC 27599, USA
          \and
             Astrophysics Research Centre, School of Mathematics and Physics, Queen’s University Belfast, BT7 1NN, UK
          \and
             Physics Department and Tsinghua Center for Astrophysics (THCA), Tsinghua University, Beijing 100084, China
          \and
             Department of Physics and Astronomy, Aarhus University, Ny Munkegade 120, 8000 Aarhus C, Denmark
          \and
             Instituto de Astrof\'isica de Canarias, E-38205 La Laguna, Tenerife, Spain
          \and
             Departamento de Astrof\'isica, Universidad de La Laguna, E-38206 La Laguna, Tenerife, Spain
          \and
             Department of Physics and Astronomy, University of Turku, FI-20014 Turku, Finland
          \and
             The Oskar Klein Centre, Department of Astronomy, Stockholm University, AlbaNova, SE-10691 Stockholm, Sweden
           \and
             Department of Applied Physics, University of C\'adiz, Campus of Puerto Real, 11510 C\'adiz, Spain
         \and
             Finnish Centre for Astronomy with ESO (FINCA), Quantum, Vesilinnantie 5, University of Turku, 20014 Turku, Finland
          \and
           Institute for Astronomy, 2680 Woodlawn Drive, Honolulu, HI 96822-1897, USA
          \and
          Space Telescope Science Institute, 3700 San Martin Drive, Baltimore, MD 21218, USA
          \and
          Department of Physics and Astronomy, Johns Hopkins University, Baltimore, MD 21218, USA
}

   \date{Received Month 11, 2020; accepted Month xx, 202x}

  \abstract
 {We present the results of our monitoring campaigns of the luminous red novae (LRNe) AT~2020hat in NGC~5068 and AT~2020kog in NGC~6106.
The two objects were imaged (and detected) before their discovery by routine survey operations. They show a general trend of slow luminosity rise, 
lasting at least a few months. The subsequent major LRN outbursts were extensively followed in photometry and spectroscopy. The light curves present 
an initial short-duration peak, followed by  a redder plateau phase. AT~2020kog is a moderately luminous event peaking at $\sim7\times10^{40}$ erg s$^{-1}$, 
while AT~2020hat is almost one order of magnitude fainter than AT~2020kog, although it is still more luminous than V838~Mon.
 In analogy with other LRNe, the spectra of AT~2020kog change significantly with time. They resemble those of type IIn supernovae at early phases, 
then they become similar to those of K-type stars during the plateau, and to M-type stars at very late phases. In contrast, AT~2020hat already 
shows a redder continuum at early epochs, and its spectrum shows the late appearance of molecular bands.
A moderate-resolution spectrum of AT~2020hat taken at +37~d after maximum shows a forest of narrow P~Cygni lines of metals with velocities 
of 180 km s$^{-1}$, along with an H$\alpha$ emission with a full-width at half-maximum velocity of 250 km s$^{-1}$. For AT~2020hat, 
a robust constraint on its quiescent progenitor is provided by archival images of the Hubble Space Telescope. The progenitor is clearly 
detected as a mid-K type star, with an absolute magnitude of $M_{F606W}=-3.33\pm0.09$ mag and a colour of $F606W-F814W=1.14\pm0.05$ mag, which are 
inconsistent with the expectations from a massive star that could later produce a core-collapse supernova.
Although quite peculiar, the two objects nicely match the progenitor versus light curve absolute magnitude correlations discussed in the literature. }


   \maketitle
%

\section{Introduction}

Luminous red novae (LRNe) form a novel class of stellar transients with luminosities intermediate between novae and core-collapse supernovae (SNe).
These gap transients \citep{kas12,pasto19} have a characteristic double-peak light curve, early spectra similar to those of type IIn SNe, and late spectra
that migrate from matching those of intermediate stellar types to those of mid-to-late M-type stars \citep[see,][and references therein]{pas19a}. These properties
allow us to securely discriminate LRNe from the so-called intermediate-luminosity red transients \citep[ILRTs\footnote{ILRTs always show a very slow spectral evolution and
prominent Ca lines, in particular the typical narrow [Ca~II] $\lambda\lambda$ 7291, 7324 doublet in emission.}, e.g.][]{bot09,bon09,cai18,max20a}, which are believed 
to originate from eruptions or terminal explosions of super-asymptotic giant branch (S-AGB) stars. Nonetheless, some observational overlap between the two classes has 
been occasionally noted, giving rise to controversial cases \citep[e.g. M85-2006OT1 and AT~2018hso;][]{kul07,pas07,cai19}. 

The LRN phenomenon can be comfortably explained in terms of post common envelope evolution and eventually coalescence in binary systems whose components span a wide range 
of masses. However, the physical processes leading to the common envelope ejection and the path to coalescence are still debated \cite[e.g.][]{iva13,pej17,met17,sag19,mac18,sok20a,sok20b}.
The characterisation of this species of gap transients is also necessary  to provide reliable estimates of their rates, both in the Galaxy and in the local Universe, although it is now 
clear that the frequency of LRNe significantly depends on their luminosity and the total mass of the system \citep{koc14,how20}. Finally, the study of the evolution of massive binaries 
is a hot topic, as this has important implications in the formation of close compact binaries \citep{kle20}, which are sources of gravitational waves \citep{abb16,abb17}.

Collecting good datasets covering all the crucial phases of the LRN evolution is a necessary condition to develop reliable theoretical models \citep[see, e.g.][]{pej16a,pej16b,met17,sag19,mac20a,mac20b}.
However, so far  only a limited number of LRNe are available in the literature with extensive light curves, including pre-outburst detections and well-sampled spectral sequences.
For this reason, within the NOT Unbiased Transient Survey~2 \citep[NUTS2\footnote{\url{http://nuts.sn.ie}};][]{hol19} collaboration, we are leading a programme of the systematic 
follow-up of LRNe  \citep[e.g.][]{cai19,pas19a,pas19b}. 

In this paper, we present the outcomes of our monitoring campaigns of two LRNe in the nearby Universe: AT~2020hat and AT~2020kog. These two objects, along 
with AT~2019zhd in M~31 \citep[whose study is presented in a companion paper,][]{pasto20}, were discovered in the observing semester from December 2019 to May 2020. 
Our team made a substantial effort to follow these three objects and provided excellent datasets for all of them, although the monitoring campaigns were severely complicated 
by the fact that many observational facilities had to suspend operations during the global lock-down resulting from the Covid-19 pandemic emergency.

The paper is structured as follows: Sect. \ref{Sect:hosts} reports information on the discovery, classification, host-galaxy properties, distance, and reddening 
of AT~2020hat and AT~2020kog; their light curves and spectral evolution are presented in Sect. \ref{Sect:lightcurves} and Sect. \ref{Sect:spectra}, 
respectively;  archive photometry of the quiescent progenitor system of AT~2020hat is analysed in Sect. \ref{Sect:progenitor}; the evolution of the bolometric 
luminosity, temperature, and radius is illustrated in Sect. \ref{Sect:bolom}. A general discussion on the physical properties of LRNe and final remarks follow in Sect. \ref{Sect:discussion}. 

   \begin{figure}
   \centering
   {\includegraphics[angle=0,width=8.9cm]{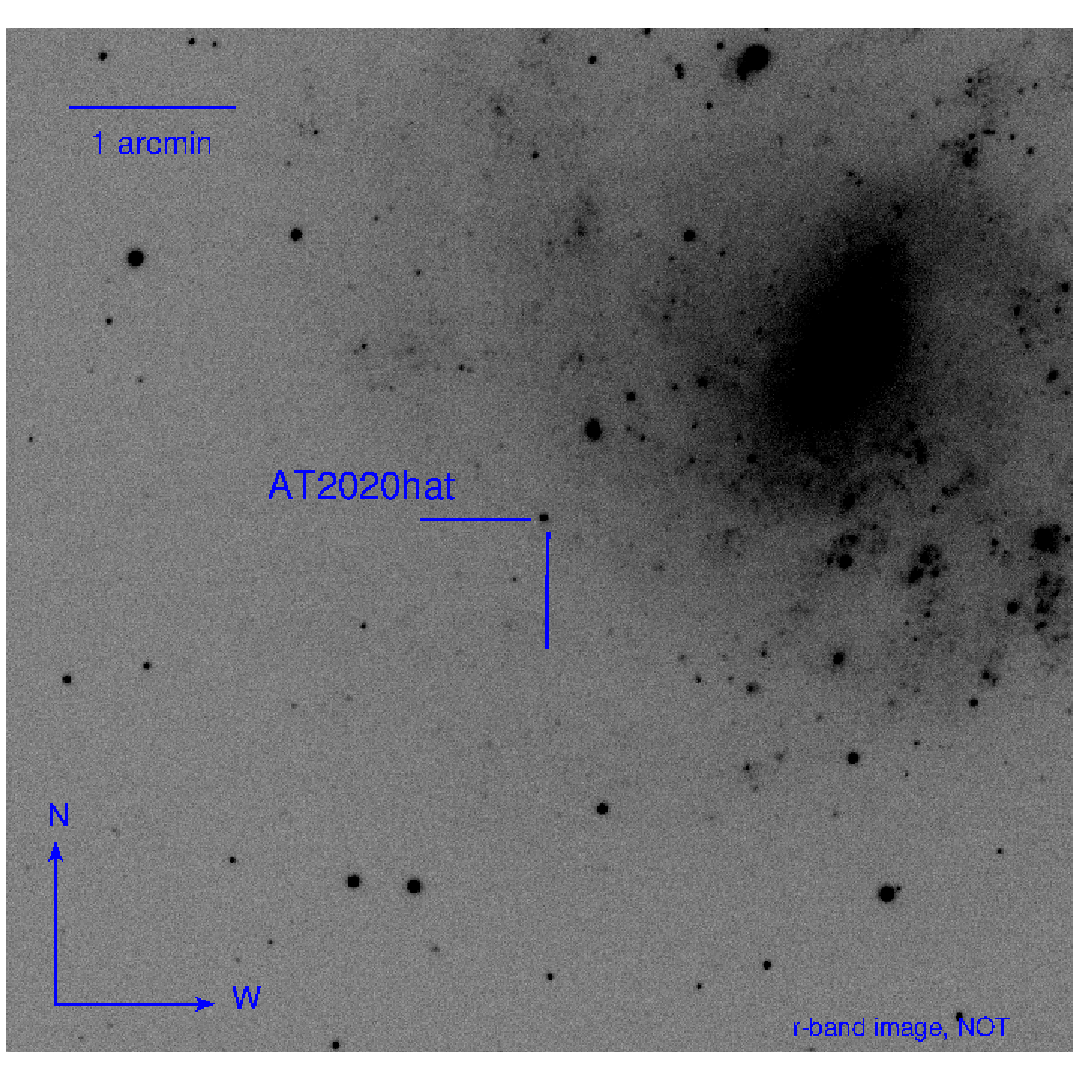}
\includegraphics[angle=0,width=8.9cm]{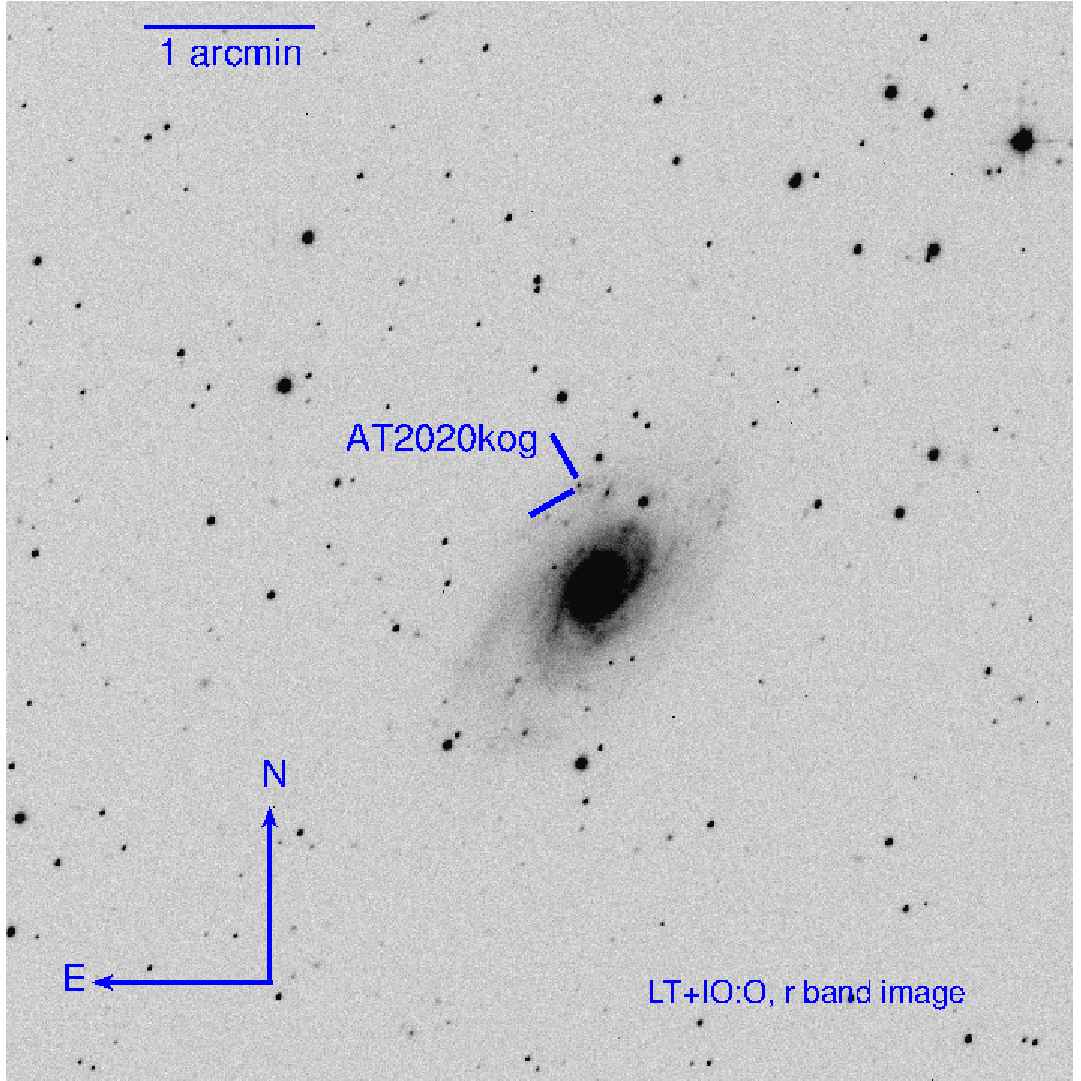}}
      \caption{Finding charts of AT~2020hat ({\sl top panel}) and AT~2020kog ({\sl bottom panel}), showing the host environments.
              }
         \label{Fig:findingcharts}
   \end{figure}
%

\section{AT~2020hat and AT~2020kog: Discoveries and host galaxies} \label{Sect:hosts}

%
   \begin{figure*}
   \centering
   {\includegraphics[angle=0,width=9.1cm]{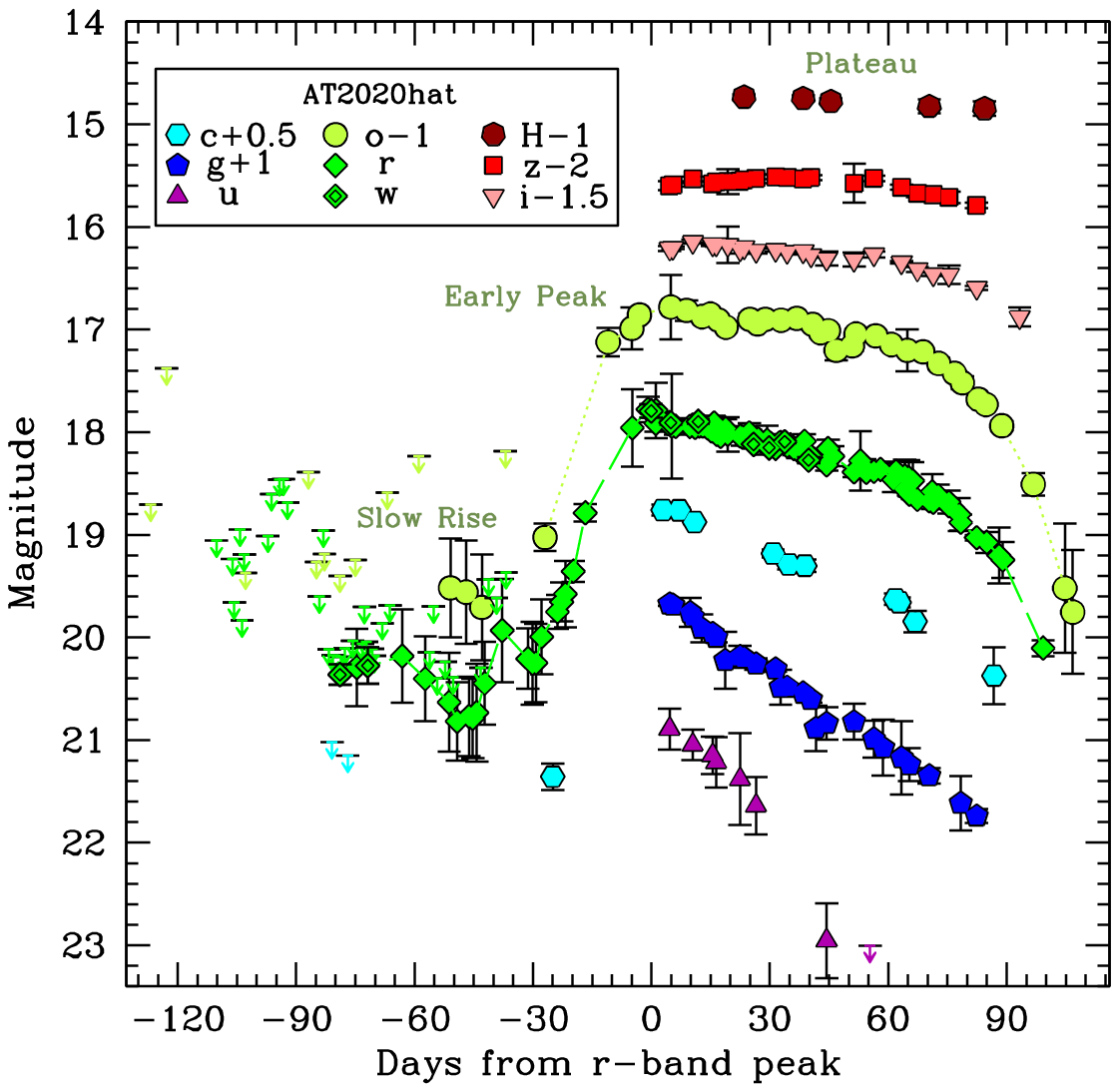}
\includegraphics[angle=0,width=9.2cm]{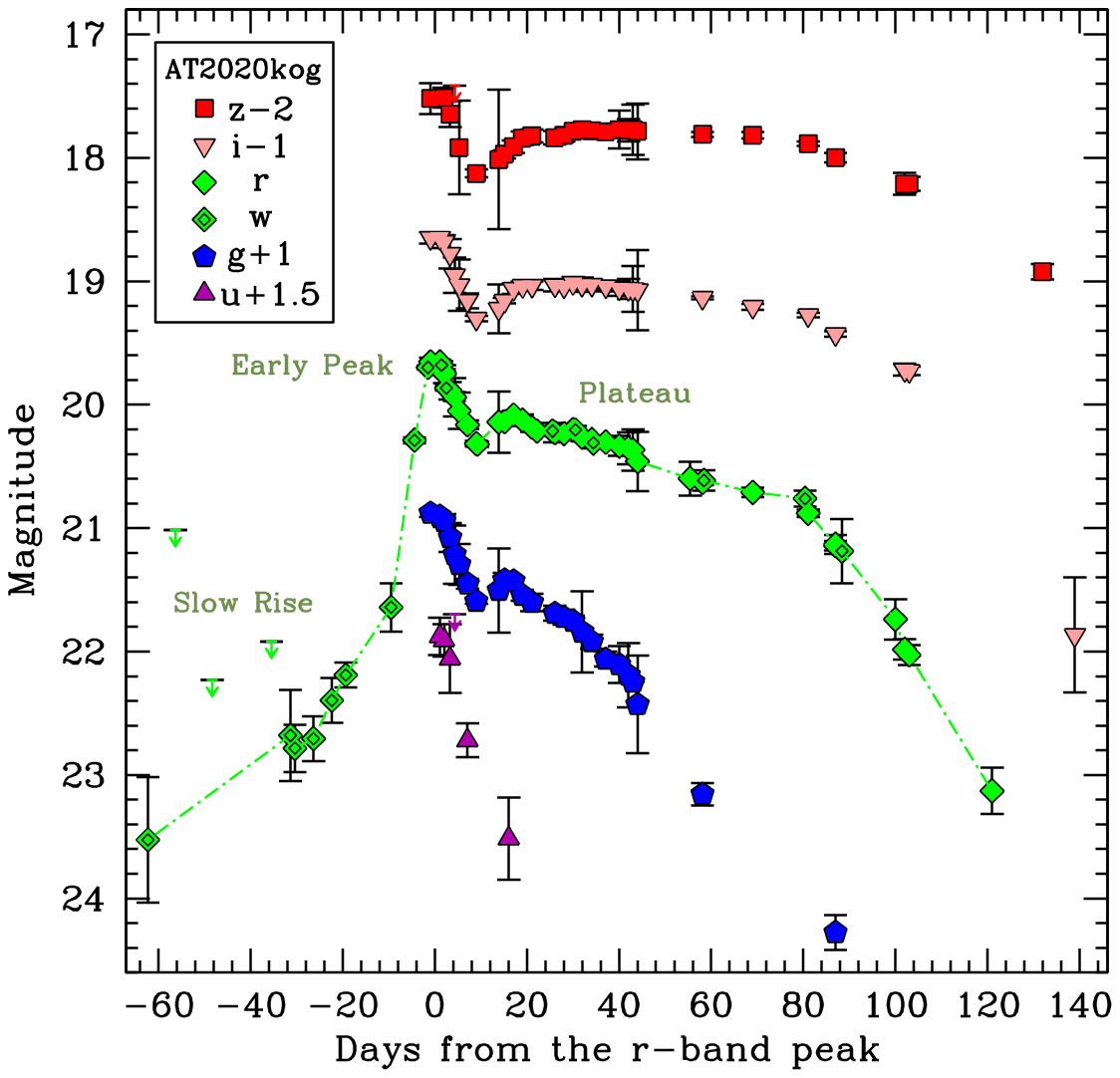}}
      \caption{Multi-band light curves of AT~2020hat ({\sl left panel}) and AT~2020kog ({\sl right panel}), including the data from the surveys.
              }
         \label{Fig:lightcurves}
   \end{figure*}
%

       \begin{table*}
      \caption[]{\label{tab:TabNIR} Near-infrared $JHK$ (Vega mag) photometry of AT~2020hat.}         
      $$     \begin{tabular}{cccccc}
            \hline\hline
            \noalign{\smallskip}
            Date &  MJD & $J$ & $H$ & $K$ & Instrument \\ \hline                    
2020-05-08 & 58977.96 &  --            & 15.736 (0.029) &  --            &  2.0m~LT + IO:I \\
2020-05-23 & 58992.97 &  --            & 15.748 (0.029) &  --            &  2.0m~LT + IO:I \\
2020-05-30 & 58999.95 & 16.320 (0.073) & 15.777 (0.028) & 15.681 (0.042) &  2.56m~NOT + NOTCam \\
2020-06-24 & 59024.91 & 16.369 (0.044) & 15.825 (0.067) & 15.674 (0.027) &  2.56m~NOT + NOTCam \\
2020-07-08 & 59038.90 &  --            & 15.846 (0.071) &  --            &  2.56m~NOT + NOTCam \\

        \noalign{\smallskip}
            \hline
         \end{tabular}
$$
   \end{table*}

AT~2020hat\footnote{The transient is also known with the survey names ATLAS20jxk, PS20cjf, and ZTF20aawdwch.}  was discovered on 2020 April 12.48 UT in the Sc-type 
galaxy NGC~5068 by the Asteroid Terrestrial-impact Last Alert System \citep[ATLAS,][]{ton18,smi20}, at an ATLAS orange-band magnitude of $o = 17.83\pm0.08$ mag. 
The coordinates of the object are $RA = 13^{h}19^{m}01\fs927$, $Dec = -21\degr03\arcmin16\farcs37$ (equinox J 2000.0), which is offset by $55\farcs4$ south and $99\farcs6$ 
east from the centre of the host galaxy (see Fig. \ref{Fig:findingcharts}, top panel). The object was classified as an LRN after maximum light by \citet{reg20}. The transient 
was also observed in the radio domain, but no counterpart was found down to 3$\sigma$ limits of about 0.06 mJy at 5.5 GHz and 0.04 mJy at 9.0 GHz \citep{ryd20}. 

The host, NGC 5068, is a nearby face-on Sc-type galaxy\footnote{According to HyperLeda; \url{http://leda.univ-lyon1.fr/}.} with a redshift of $z=0.002235\pm0.000003$ \citep{pis11}. 
Given its proximity, the redshift does not provide a good constraint on the galaxy distance. For this reason, in this paper we adopt the distance estimate obtained through the tip 
of red-giant branch method, which provides a distance of $d=5.16\pm0.21$ Mpc \citep{kar17}, hence  a distance modulus of $\mu = 28.56 \pm 0.08$ mag. We note that this value is 
remarkably similar to another estimate based on the planetary nebulae luminosity function \citep[$\mu = 28.68 \pm 0.08$ mag,][]{her08}.

The reddening towards AT~2020hat is dominated by the Milky Way component, $E(B-V)_{MW} = 0.09$ mag \citep{sch11}. In fact, the analysis of the spectra of the transient 
(as we see in Sect. \ref{Sect:spectra}) does not indicate additional host galaxy reddening, as expected considering the peripheral location of the object in its host. 

AT~2020kog, also known as PS20dgq, was discovered by the Panoramic Survey Telescope and Rapid Response System (Pan-STARRS) Search for Kilonovae \citep{sma19} on 2020 May 18.49 UT, in an 
outer spiral arm of the Sc-type galaxy NGC~6106. The coordinates of the object are $RA = 16^{h}18^{m}47\fs652$, $Dec = +07\degr25\arcmin16\farcs86$ (equinox J 2000.0), with an offset of 
$42\farcs7$ north and $9\farcs1$ east from the core of the host galaxy (Fig. \ref{Fig:findingcharts}, bottom panel). The survey discovery magnitude was $w=20.30\pm0.10$ mag \citep{ful20}.

The redshift of NGC~6106 is $z=0.00483\pm0.00002$. Adopting a standard cosmology with $H_0=73$ km s$^{-1}$ Mpc$^{-1}$, $\Omega_{matter}$ = 0.27, and $\Omega_{vacuum}$ = 0.73, 
a distance of $23.8\pm1.7$ Mpc (corrected for Virgo Infall)\footnote{We note that other kinematic distance estimates span a relatively wide range of values, from 20.7 to 26.1 Mpc.} 
is obtained \citep{mou00}, which provides  $\mu = 31.89 \pm 0.15$ mag. Numerous distance estimates have been published using the Tully-Fisher method, spanning a wide range of inferred 
values (from about 21 to 28 Mpc). Here we adopt the most recent estimate from \citet{tul16} adapted to our cosmological assumptions: $d=22.5\pm2.0$ Mpc, which gives  $\mu = 31.76 \pm 0.45$ mag. 
This distance, which is adopted hereafter in the paper, is slightly lower than the kinematic distance.

The dust extinction within the Milky Way in the direction of the transient is $E(B-V)_{MW} = 0.06$ mag \citep{sch11}. In addition, the detection of a prominent narrow interstellar feature 
of Na I $\lambda\lambda$5890,5896 (Na~ID) in the spectra of the transient suggests the existence of additional reddening within the host galaxy. With a host galaxy contribution of 
$E(B-V)_{host} = 0.31\pm0.06$ mag (see Sect. \ref{20kog_spec} for details), we find that a total colour excess due to line-of-sight dust extinction is $E(B-V)_{tot} = 0.37\pm0.07$ mag. 

\begin{table*}
\caption{\label{tab:speclog}Log of spectroscopic observations of the two LRNe. Phases are from their $r$-band light curve peak,
that in AT~2020hat and AT~2020kog are at $MJD_{r,max} = 58954.0\pm1.5$ and $MJD_{r,max} = 58991.9\pm0.7$, respectively.}
\centering
\begin{tabular}{lcccccc}
\hline\hline
Date&MJD&Phase&Instrumental configuration& Exptime (s) & Range (nm) & Resolution (nm) \\
\hline
\multicolumn{7}{c}{AT~2020hat} \\
\hline
2020-04-16 & 58955.99 &  +2.0 & NOT+ALFOSC+gm4 & 3600 & 340-960 & 1.4 \\
2020-04-21 & 58960.01 &  +6.0 & NOT+ALFOSC+gm4 & 3600 & 340-960 & 1.8 \\
2020-04-28$^\star$ & 58967.99 & +14.0 & LT+SPRAT+VPH600 & 2100 & 400-800 & 1.4 \\
2020-05-02 & 58971.07 & +17.1 & TNG+DOLORES+LRB & 2700 & 320-795 & 1.4 \\
2020-05-03 & 58972.02 & +18.0 & TNG+DOLORES+LRR & 2700 & 500-1030 & 1.3 \\
2020-05-17 & 58986.95 & +33.0 & GTC+OSIRIS+R1000B+R1000R & 2400,1800 & 360-1035 & 0.7,0.8 \\
2020-05-17 & 58986.99 & +33.0 & GTC+OSIRIS+R2500R & 3000 & 559-768 & 0.34 \\
2020-05-29 & 58998.96 & +45.0 & TNG+DOLORES+LRB+LRR & 2400,2400 & 330-1035 & 1.0,0.9\\
2020-06-29 & 59029.94 & +75.9 & TNG+DOLORES+LRR & 3600 & 500-970 & 1.3 \\
\hline
\multicolumn{7}{c}{AT~2020kog} \\
\hline
2020-05-27 & 58996.18 &  +4.3 & GTC+OSIRIS+R1000B & 1500 & 365-785 & 0.7 \\
2020-06-01 & 59001.17 &  +9.3 & GTC+OSIRIS+R1000R & 3000 & 510-1035 & 0.8 \\
2020-06-14 & 59014.09 & +22.2 & GTC+OSIRIS+R1000B & 3000 & 365-785 & 0.7 \\
2020-06-29 & 59029.04 & +37.1 & GTC+OSIRIS+R1000R & 3000 & 510-1035 & 0.8 \\
2020-08-17 & 59078.91 & +87.0 & GTC+OSIRIS+R1000R & 3000 & 510-1035 & 0.8 \\
2020-08-17 & 59078.97 & +87.1 & NOT+ALFOSC+gm4 & 5400 & 390-960 & 1.4 \\
2020-08-30 & 59091.92 & +100.0 & GTC+OSIRIS+R1000R & 3000 & 510-1040 & 0.8 \\
\hline
\end{tabular}
\tablefoot{$^\star$ Low S/N spectrum, not shown in Fig. \ref{Fig:specseq}.}
\end{table*}

\section{Light curves of AT~2020hat and AT~2020kog} \label{Sect:lightcurves}

The photometric follow-up campaigns of AT~2020hat and AT~2020kog were performed after their discovery using a number of instruments available to our collaboration.
The details of the instrumental configurations are provided in the photometry tables (available in electronic form at the CDS).

Due to its faint apparent magnitude,  AT~2020kog was not detected by most of the surveys. However, important early data have been obtained by Pan-STARRS \citep[][]{cha19,fle20,mag20}, with the wide 
$w$-band filter. In contrast, AT~2020hat had a brighter apparent magnitude, and it was routinely observed by numerous surveys, including the $D < 40$ Mpc SN survey \citep[DLT40;][unfiltered data]{tar18}, 
the Zwicky Transient Facility \citep[ZTF;][in the Sloan-$g$ and $r$ bands]{mas18}\footnote{ZTF forced photometry is released through the Lasair (\url{https://lasair.roe.ac.uk/}) and ALeRCE 
(\url{https://alerce.online/}) brokers.}, the All-Sky Automated Survey for Supernovae \citep[ASAS-SN;][in the $g$ band]{sha14,koc17}\footnote{ASAS-SN photometry is publicly released through the Sky Patrol 
ASAS-SN interface (\url{https://asas-sn.osu.edu}).}, the  Asteroid Terrestrial-impact Last Alert System project \citep[ATLAS:][in the ATLAS-cyan and orange bands]{ton18} and Pan-STARRS (in the $i$ and $z$ bands). 
We remark that in our paper the unfiltered DLT40 data were calibrated to Sloan-$r$ magnitudes, while the observations with ATLAS-orange ($o$) and ATLAS-cyan ($c$) filters were left in their original photometric 
systems. Finally, although no transformation was applied to the Pan-STARRS-$w$ band magnitudes provided by the survey, these data match our  Sloan-$r$ observations very well.

Our imaging data needed some preliminary processing, including bias, flat-field, and (when useful) fringing pattern corrections. Near-infrared (NIR) images required additional preliminary steps, including 
building a sky image for each filter obtained from median-combining dithered science frames. Then, the sky image was subtracted from individual science frames, which were finally combined to increase the 
signal-to-noise ratio (S/N). Optical and NIR imaging data were reduced using the {\sc python/pyraf} {\sc SNOoPY} pipeline \citep{cap14}. The software allowed us to perform astrometric calibration of the images, 
PSF-fitting photometry of the transient (with or without template subtraction)\footnote{AT~2020kog was located in a crowded region of its host galaxy, hence we subtracted  template images of NGC~6106 obtained 
from the Pan-STARRS survey ($griz$ filters) and SDSS ($u$ filter). In contrast, AT~2020hat was in a clean and peripheral location of the host galaxy, hence template subtraction was not necessary.}, and the 
subsequent photometric calibration making use of catalogues of reference stars. For the optical Sloan band images, we used the Sloan catalogue, while for the NIR data we used the Two Micron All-Sky Survey (2MASS) 
catalogue \citep{skr06}. The unfiltered DLT40 photometry was calibrated with reference to the Sloan-$r$ photometry.

The final light curves of the two objects are shown in Fig. \ref{Fig:lightcurves}, while the optical photometry Tables E1 and E2 are made available in electronic form at the CDS.
The NIR observations of AT~2020hat are instead reported in Table \ref{tab:TabNIR}.

The field of AT~2020hat was routinely sampled before the LRN discovery by public surveys. These observations are crucial to build the light curve before the LRN outburst (Fig. \ref{Fig:lightcurves}, left panel). 
In particular, a faint source at the position of the transient is observed starting about three months before the AT~2020hat peak luminosity. In the pre-outburst phase, the light curve shows some fluctuations over 
a general brightening trend. From about one month before maximum, the best-sampled $r$-band light curve shows a much more evident luminosity rise of about 2.5 mag, reaching the peak at $r_{max} = 17.79\pm0.02$ mag 
(on $MJD_{r,max} = 58954.0\pm1.5$)\footnote{The time of the $r$-band first light curve peak is used as reference epoch throughout the paper for both AT~2020hat and AT~2020kog.}. The low-contrast light-curve peak 
is followed by a nearly linear decline in the blue bands, while a sort of plateau lasting about 70-80~d is observed in the red bands. This phase is followed by a rapid decline that unfortunately was not sampled 
well due to the modest visibility from our northern facilities. The decline rates in the different bands between the maximum and $\sim+70$~d are as follows: $\gamma_u = 5.22\pm0.61$, $\gamma_g = 2.59\pm0.07$, 
$\gamma_r = 0.97\pm0.03$, $\gamma_o = 0.61\pm0.06$,  $\gamma_i = 0.43\pm0.03$, and $\gamma_z = 0.04\pm0.05$ mag~(100~d)$^{-1}$ . The late-time ($>$80~d) decline rates in the Sloan-$r$ and ATLAS-$o$ bands steepen 
to $\gamma_r = 6.7\pm0.9$ mag~(100 d)$^{-1}$ and $\gamma_o = 9.4\pm0.8$ mag~(100~d)$^{-1}$, respectively.

A few observations of AT~2020hat were also obtained in the NIR domain by the 2.0~m Liverpool Telescope (LT) with IO:I and the 2.56~m Nordic Optical Telescope (NOT) with NOTCam. While only two epochs were obtained in the 
$J$ and $K$ bands, observations in the $H$ band were obtained at five epochs during the plateau, with the object staying at roughly constant luminosity, around $H \sim 15.8$ mag (Fig. \ref{Fig:lightcurves}, left panel).

AT~2020kog was detected by Pan-STARRS a couple of months before the outburst peak (see Fig. \ref{Fig:lightcurves}, right panel). During the pre-outburst phase, until one month before maximum, the $r$-band light 
curve initially rises with a rate of 2.4$\pm$0.3 mag (100~d)$^{-1}$. The $r$-band light curve brightening becomes faster in the following two weeks, with a rate of 5.8$\pm$0.2 mag (100~d)$^{-1}$, to finally 
increase by further $\sim$2 mag in about 10 days by the time of the first blue peak. The epoch of the maximum and the peak magnitude of the $r$-band light curve of AT~2020kog can be inferred from a low-order 
polynomial fit: We obtain $MJD_{r,max} = 58991.9\pm0.7$ and $r_{max}=19.48\pm$0.02 mag. Pre-maximum information is not available for the $g$ band, so we adopt $MJD_{g,max} \sim58991.0$ and $g_{max}\sim19.88$ mag 
as indicative parameters for the $g$-band maximum.

   \begin{figure*}
   \centering
   {\includegraphics[angle=270,width=15.4cm]{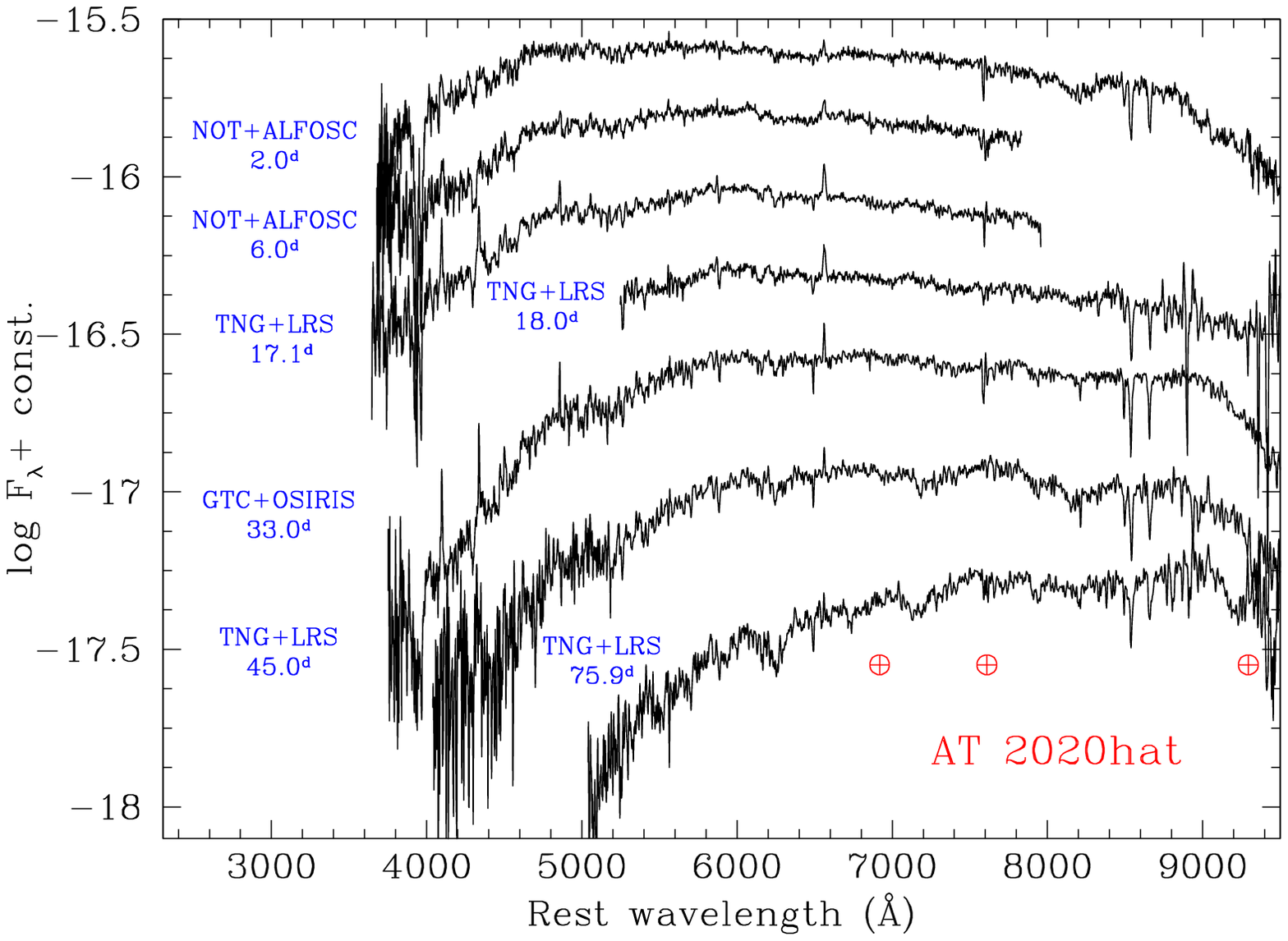}
\includegraphics[angle=270,width=15.4cm]{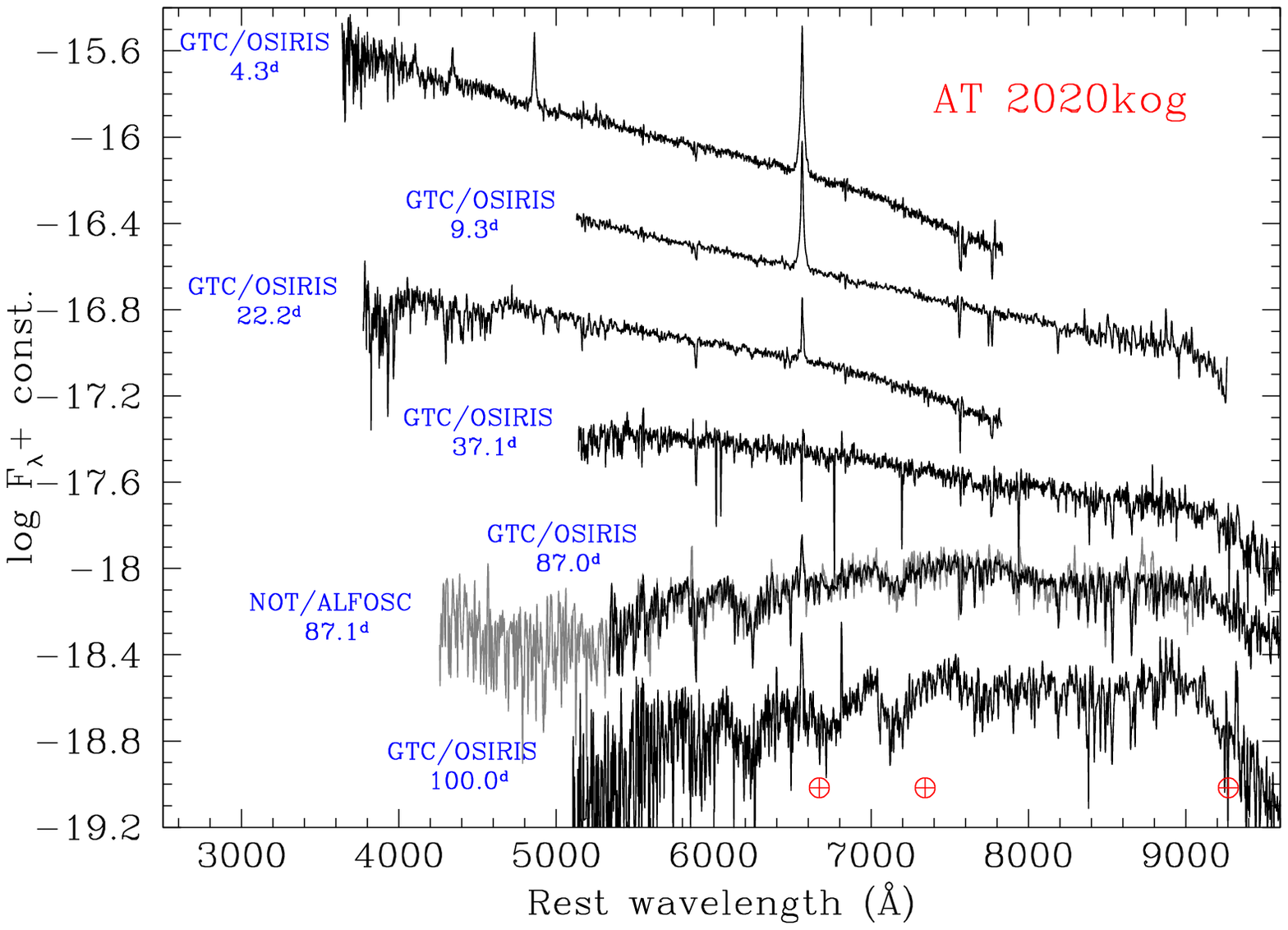}}
      \caption{Spectroscopic evolution of AT~2020hat ({\sl top panel}) and AT~2020kog ({\sl bottom panel}). The spectra are redshift and reddening corrected. The phases reported on the left
are with respect to the  light curve maximum.
              }
         \label{Fig:specseq}
   \end{figure*}
%

After the peak, the $r$-band light curve fades with a slope of $\gamma_r =8.2\pm0.2$ mag~(100~d)$^{-1}$ and reaches a local minimum of $r \sim 20.3$ mag at about 10~d past-maximum, 
followed by a modest re-brightening ($r \sim 20.1$ mag). From three weeks after maximum, the light curve declines almost linearly, with a rate of $\gamma_r = 1.3\pm0.1$ mag (100~d)$^{-1}$. 
A much larger steepening is observed from about 80~d after maximum in all bands, although the decline rates are less steep in the Sloan-$i$ and $z$ bands. At phases $>$ 80~d, we find 
$\gamma_r =5.7\pm0.2$, $\gamma_i =4.6\pm0.7$, and $\gamma_z =2.0\pm0.2$ mag~(100~d)$^{-1}$.

\section{Spectroscopic data of AT~2020hat and AT~2020kog} \label{Sect:spectra}

Spectroscopic observations of the two objects were performed using the 10.4~m Gran Telescopio Canarias (GTC) equipped with OSIRIS, the 3.58~m Telescopio Nazionale Galileo (TNG) equipped 
with Dolores (LRS), the NOT with ALFOSC, and the 2.0~m LT with SPRAT. All these telescopes are hosted at the Observatorio del Roque de los Muchachos in La Palma (Canary Islands, Spain). 
Information on the instrumental configurations used for the spectroscopic data is given in Table \ref{tab:speclog}.

The spectra were reduced using standard tasks in {\sc IRAF}\footnote{IRAF is distributed by the National Optical Astronomy Observatory, which is operated by the Association of Universities 
for Research in Astronomy (AURA) under a cooperative agreement with the National Science Foundation.}, with the exception of the ALFOSC data which were reduced using the {\sc Python}-based 
{\sc FOSCGUI} pipeline\footnote{\url{https://sngroup.oapd.inaf.it/foscgui.html}} developed by E. Cappellaro. After the traditional bias and flat-field correction of the science frames, 1-d 
spectra were optimally extracted and wavelength calibrated using arc-lamp spectra. The accuracy of the wavelength calibration was checked by measuring the wavelengths of the [O~I] night sky 
lines. Then, the spectra were flux calibrated using spectrophotometric standards taken during the same night as the LRN observation. The flux calibration was finally fine-tuned using the 
photometric information, and the telluric absorption bands of O$_2$ and H$_2$O were removed using the spectra of hot standard stars.

AT~2020hat was discovered quite late, and we could not follow it spectroscopically during the pre-maximum evolution. Nonetheless, the observational cadence after discovery is excellent and 
spans over 100 days of its evolution. The observational campaign of AT~2020kog is more complete, although it was hampered by the faint apparent magnitude of the transient and, because of 
its northern declination, the lack of support of southern facilities. As a consequence, the object was only observed with GTC and NOT, starting from soon after the blue peak to +100 d.
Despite the limited number of spectra, we have been able to sample all of the crucial phases of the evolution of AT~2020kog. The two spectral sequences are shown in Fig. \ref{Fig:specseq}.

\subsection{AT~2020hat spectra}

The spectra of AT~2020hat are typical of the cooler, red-peak (or plateau) phase of LRNe. This either implies that we failed to see an earlier hotter phase, which is unlikely, or that this LRN did not 
become very hot at the early phases of the outburst, as occasionally observed in lower luminosity Galactic LRNe such as V1309Sco \citep{mas10,tyl11} and V838~Mon \citep{mun02,gor02,kim02,cra03,tyl05}.

The spectra show weak Balmer lines with P Cygni profiles and a forest of narrow absorptions due to metal lines. These features are responsible for line blanketing at the blue wavelengths, although many 
multiplets are clearly detected at longer wavelengths. It is important to note that Ca~II H$\&$K and Ca~II NIR triplet are among the most prominent absorption lines in the spectra of AT~2020hat.
We also possibly identify O~I $\lambda\lambda$7772,7777, although O~I $\lambda$8446 is not unequivocally seen. 

   \begin{figure}
   \centering
   \includegraphics[width=9.2cm,angle=0]{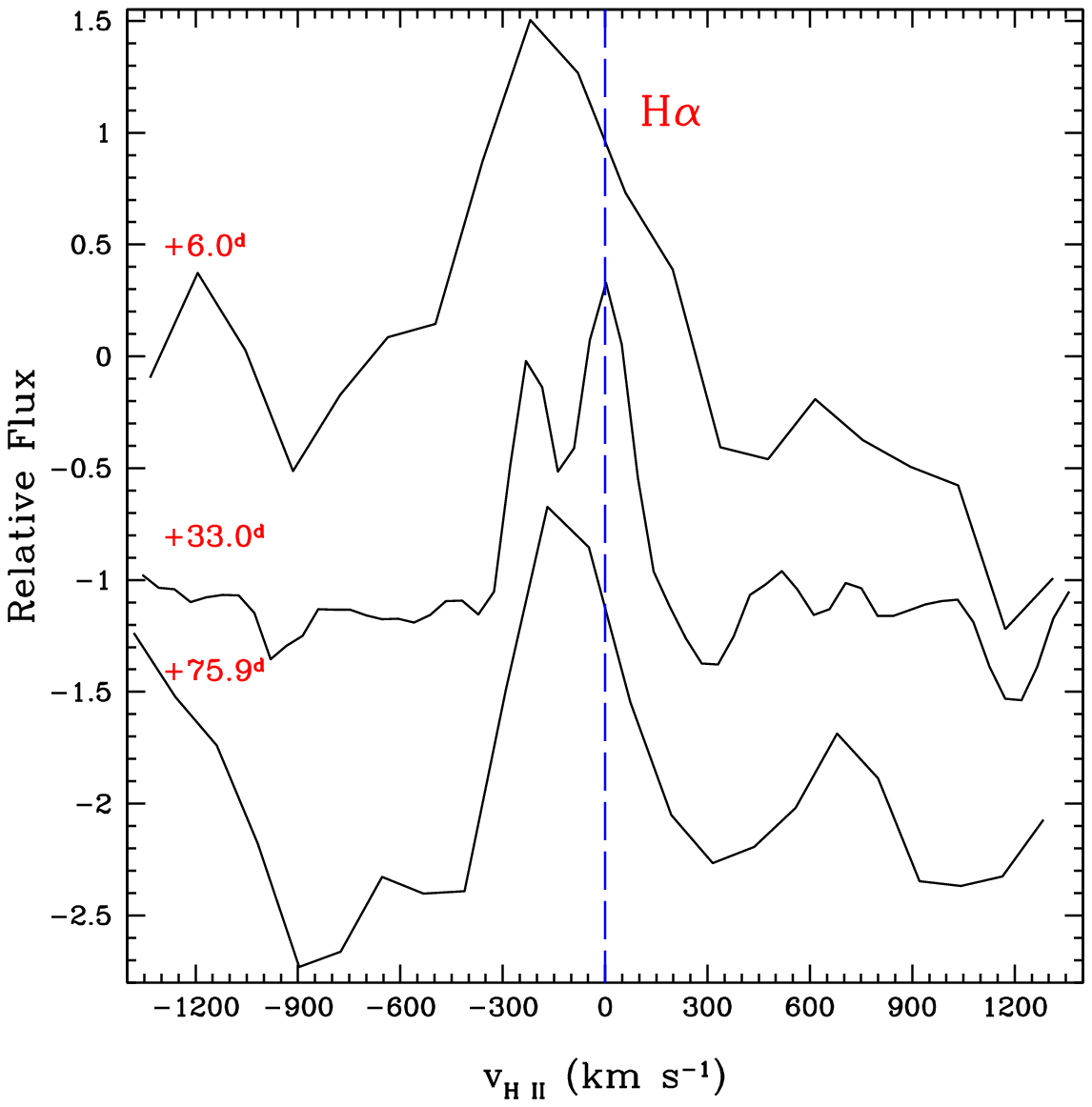}
   \caption{ \label{fig:Halpha} Blow-up of the H$\alpha$ region in the mid-resolution OSIRIS spectrum of AT~2020hat at phase +33~d, compared with two low-resolution spectra taken at earlier (+6~d) and 
later (+75.9~d) phases. As no H~II regions in the proximity of AT~2020hat have been spectroscopically observed in order to more accurately estimate the redshift at the LRN location, the spectra have 
been corrected for the average redshift of NGC~5068 ($z=0.002235$).}
    \end{figure}

   \begin{figure*}[ht!]
   \centering
   {\includegraphics[angle=270,width=14.6cm]{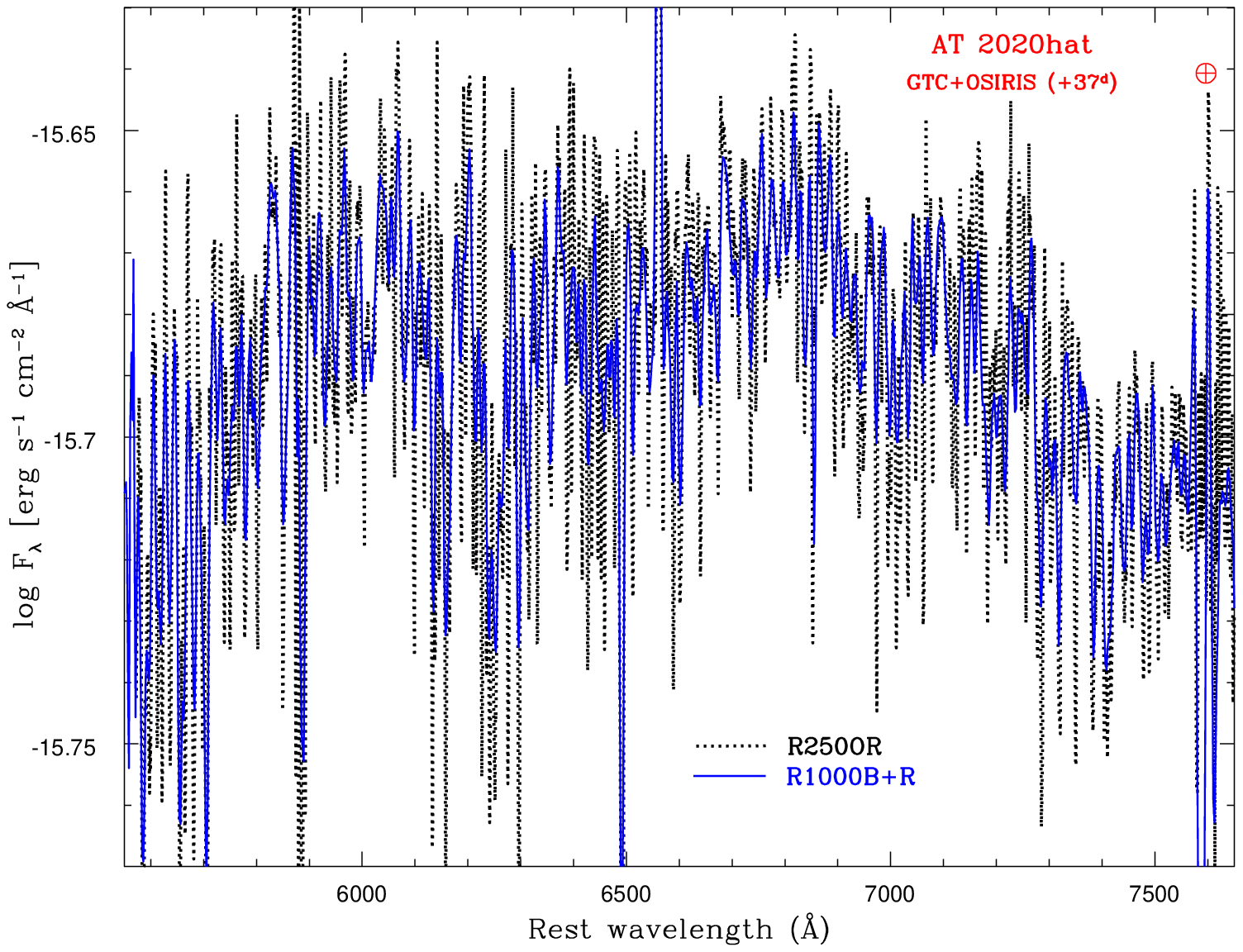}
\includegraphics[angle=270,width=14.6cm]{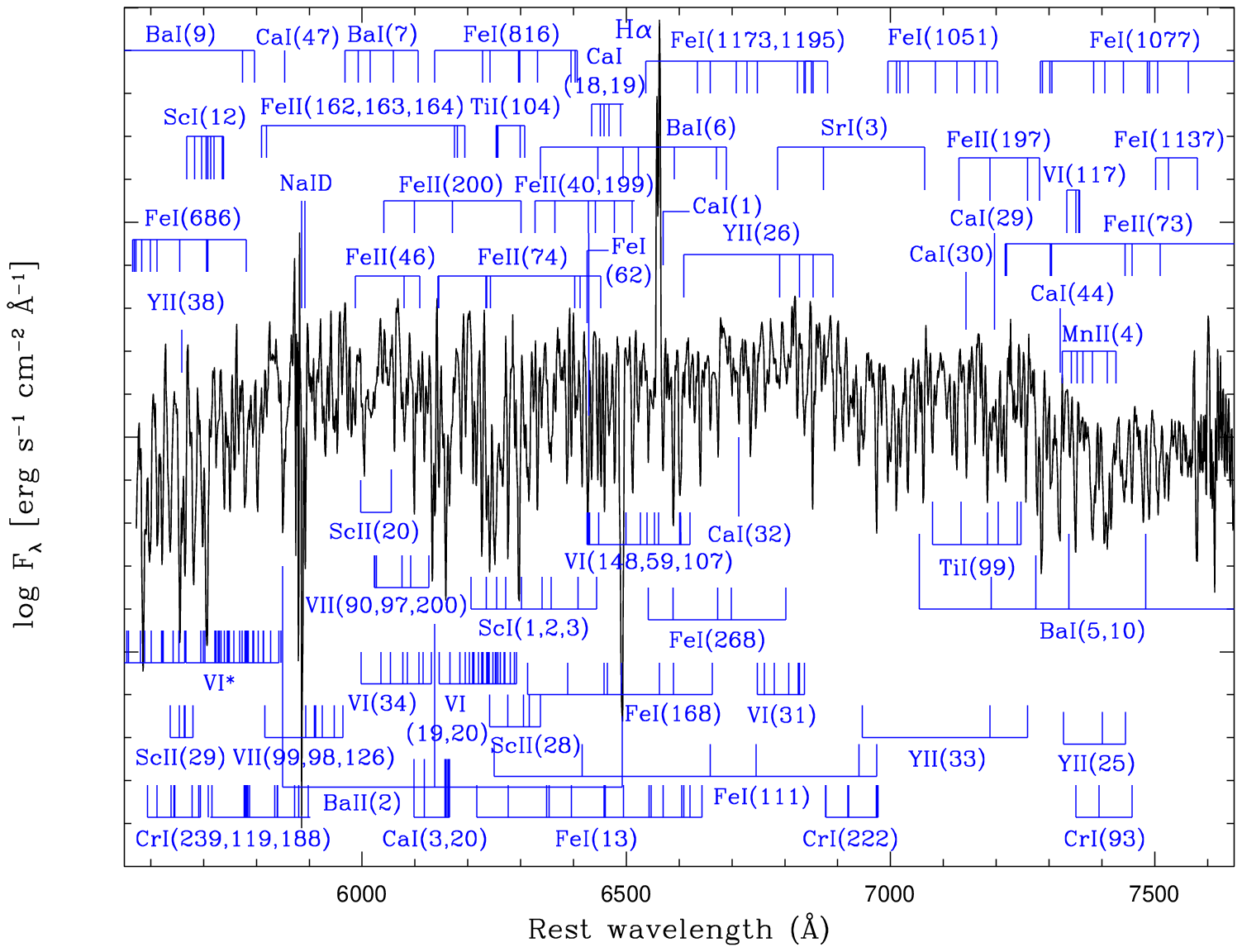}}
      \caption{{\sl Top panel:} Comparison between the highest resolution (R2500R; black dotted line) and the low resolution (R1000B+R1000R; blue solid line) GTC spectra of AT~2020hat at phase +33d. 
{\sl Bottom panel:} Line identification in the R2500R GTC spectrum at phase +33d.
              }
         \label{Fig:lineid}
   \end{figure*}

A detailed line identification and more precise velocity measurements can be performed in the region 560-760 nm using the moderate-resolution GTC+OSIRIS R2500R spectrum of AT~2020hat at phase +33 d.
We note that H$\alpha$ shows a resolved broader emission component with a full-width at half-maximum (FWHM) velocity ($v_{FWHM}$) of $250\pm30$ km s$^{-1}$ (after correction for spectral resolution), 
atop of which a narrower absorption  feature is observed. Adopting a redshift of $z=0.002235$ (see Sect. \ref{Sect:hosts}), this feature is blueshifted by about 120$\pm$5 km s$^{-1}$ from the rest wavelength 
(Fig. \ref{fig:Halpha}). An indicative estimate for the photosperic velocity at +33~d can be inferred by measuring the position of the minima of the prominent Ba~II (multiplet 2) lines, which is found to be 
$175\pm10$ km s$^{-1}$. In Fig. \ref{Fig:lineid} (top panel), we compare the R2500R spectrum with the combined R1000B+R1000R spectrum obtained at the same epoch. The comparison shows that the narrow features 
observed in the R2500R spectrum match those observed in the lower resolution configuration, confirming that these are real spectral absorption lines and not noise patterns. A line identification on the R2500R 
spectrum is shown in Fig. \ref{Fig:lineid} (bottom panel). The spectrum shows a number of multiplets of metals, which were identified following \citet{moo45}, including Ca~I, Ba~I, Ba~II, Sc~I, Sc~II, Fe~I, 
Fe~II, V~I, V~II, Ti~I, Ti~II, and Na~I. We note that Y~II and Mn~II are also tentatively identified, as these ions do not generate prominent features in the AT~2020hat spectrum.

 \subsection{AT~2020kog spectra} \label{20kog_spec}

   \begin{figure*}
   \centering
   \includegraphics[angle=0,width=20.0cm]{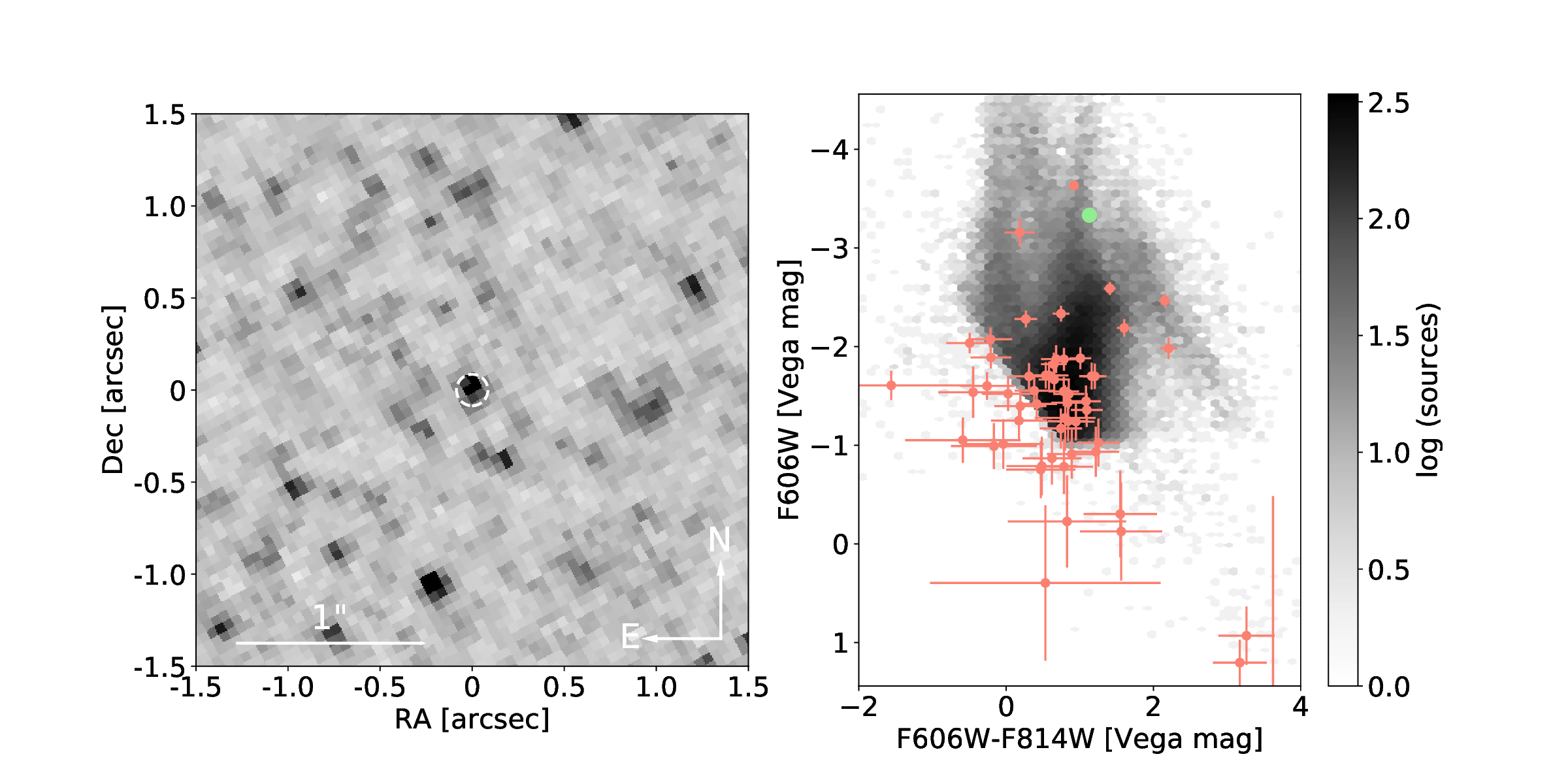}
      \caption{{\sl Left panel:} Location of AT~2020hat in the HST/ACS + $F606W$ pre-discovery image. The dashed white circle is centred on the transformed position of the transient and has a radius of 3$\times$ 
our positional uncertainty. {\sl Right panel:} Grey density plot shows the colour-magnitude diagram for all sources detected in both $F606$W and $F814W$ filters at a combined S/N$>5$. A foreground reddening $A_V=0.29$ 
mag and distance modulus $\mu=28.56$ mag have been corrected for. The sharp diagonal detection cut is due to limiting magnitude, as determined by the S/N threshold. Pink points are sources within 1\arcsec of AT 2020hat 
as indicated in the left panel, for which the S/N$>$5 detection threshold has been relaxed (as a consequence, they are plotted with large error bars). The green point is the progenitor candidate for AT~2020hat.}
         \label{Fig:progenitor}
   \end{figure*}

The first two spectra of AT~2020kog (Fig. \ref{Fig:specseq}, bottom panel) were obtained a few days after the initial blue peak. These early spectra have also been used to estimate the host galaxy 
reddening towards AT~2020kog. The narrow interstellar feature of Na~ID is well detected in these low-resolution spectra, with an equivalent width (EW) of $2.0\pm0.4$~\AA. Adopting the relation between 
EW(Na~ID) and $E(B-V)$ presented by \citet{tur03}\footnote{We remark, however, that the estimates of interstellar extinction inferred through EW(Na~ID) should be taken with caution, in particular when 
EW(Na~ID) $>1$~\AA~\citep[see discussions in][]{mun97,poz12,phi13,str18}.}, a host galaxy extinction value of  $E(B-V)_{host} = 0.31\pm0.06$ mag is inferred, hence a total colour excess of $E(B-V)=0.37\pm0.07$ mag.

After correcting for the total reddening, the continuum temperature can be inferred through a black-body fit to the continuum. We find it to be $T_{BB}=11400\pm1000$ K in the +4.3 d spectrum, and it rapidly 
declines to $T_{BB}=10200\pm1100$ K at +9.3 d, $T_{BB}=8800\pm700$ K at +22.2 d, and $T_{BB}=6600\pm1100$ K at 37.1 d. In the early spectra, along with the prominent emission lines of H, we identify P~Cygni 
lines of Ca~II, O~I, and Fe~II. In particular, the most prominent Fe~II lines are those of the multiplet 42. At  phase +4.3 d, H$\alpha$ has a Lorentzian profile with $v_{FWHM} \approx 470$ km s$^{-1}$ after 
correcting for spectral resolution and $380\pm10$ km s$^{-1}$ at +9.3 d. In our third spectrum at +22.2~d, the H$\alpha$ profile becomes asymmetric and shows two unresolved Gaussian components, with a more 
prominent emission component at the rest wavelength and a weaker emission redshifted by about +350 km s$^{-1}$. In our fourth spectrum (phase $\sim37.1$ d), H$\alpha$ is very faint, showing now a P~Cygni profile, 
whose minimum is blue-shifted by $\sim$ 220 km s$^{-1}$. The evolution of the Fe~II lines is quite modest, with an expansion velocity inferred from the P Cygni minimum, which declines from about 330 km s$^{-1}$ 
in our +4.3 d spectrum to $\sim280$ km s$^{-1}$ at 22.2 d.

Two spectra were also taken at phase $\sim$87~d, with the continuum becoming redder, with now evidence for the presence of broad absorption features due to molecules (mostly TiO), and H$\alpha$ being relatively 
prominent again. We note that these spectra are remarkably similar to the spectrum of AT~2020hat at +75.9~d (see Fig. \ref{Fig:specseq}, top panel). Both the narrow H$\alpha$ in emission and the broad 
molecular bands in absorption become much stronger in the 100~d spectrum of AT~2020kog. Among the molecular band features, we clearly identify  TiO features at about 6150-6270~\AA, 6570-6880~\AA, 7050-7270~\AA, 
7590-7860~\AA,~and above 9200~\AA. This late evolution is consistent with the expectations for an LRN \citep[e.g.][]{mar99,kam09,mas10,bar14,smi16,bla17,pas19a,pas19b,cai19,bla20}.

\section{The progenitor of AT~2020hat} \label{Sect:progenitor}

Due to its proximity, the host of AT~2020hat was frequently monitored in the past. In particular, the Hubble Space Telescope (HST) fortuitously observed NGC~5068 with the Advanced Camera for Surveys (ACS) 
Wide-Field Channel on 2017 February 11, a little over three years prior to the discovery of AT~2020hat. Two images with an exposure time of 515~s each were taken in both of the $F606W$ and $F814W$ filters. 
These data were downloaded from the Mikulski Archive for Space Telescopes\footnote{\url{http://archive.stsci.edu/}}.

In order to locate the position of AT~2020hat in these images, we obtained deep imaging from the NOT+ALFOSC on the night of 2020 June 8. To avoid saturation on the transient, we selected $15\times60$~s images 
in the Sloan-$r$ band which were subsequently co-added to give a deep image with a measured FWHM of 0.9\arcsec. We used 22 point-like sources common to the drizzled (\_drc) HST+ACS $F606W$ and NOT+ALFOSC images 
in order to derive an astrometric transformation between the two images. We find an RMS scatter of these sources of only 0.14, 0.13 ALFOSC pixels in x and y (hence $\sim28$~mas) on the transformation. After 
measuring the pixel coordinates of AT~2020hat on the ALFOSC image, we were therefore able to transform this to the ACS image with a 1$\sigma$ uncertainty of only 0.44 pixels.

A source (henceforth the progenitor candidate) is clearly visible at the transformed position in Fig. \ref{Fig:progenitor}. We used the {\sc dolphot} package \citep{dol00} to measure Vega-scale magnitudes 
of $F606W =25.48\pm0.04$ mag and $F814W=24.25\pm0.03$ mag. The source is 0.7 pixels offset from our transformed position for AT~2020hat and, as this is 1.6$\sigma$, we regard it as a strong candidate for 
the progenitor. In addition, two fainter sources were detected to the south and west which are potentially consistent with the position of AT~2020hat (Fig. \ref{Fig:progenitor}). The magnitudes and colours
of these sources are $F606W =27.42\pm0.18$, with $F606W-F814W = 0.19 \pm 0.34$ mag, and $F606W=28.51\pm0.44$ mag, with $F606W-F814W = 1.55 \pm 0.50$ mag, respectively. This would imply absolute magnitudes 
of nearly $-1$ to 0 mag at our adopted distance, and we do not consider them any further here.

For our adopted distance and extinction, the progenitor candidate has an absolute magnitude of $M_{F606W}=-3.33\pm0.09$ mag and an $F606W-F814W$ colour $=1.13 \pm 0.05$ mag, which is significantly redder 
than the detected progenitors of other LRNe \citep[see discussion in section 5 of][]{pas19a}. While, in analogy with other LRNe, these photometric parameters are most likely the integrated values of the 
binary system stellar components, the absolute magnitude is too faint to be consistent with a system dominated by a massive ($>8$ M$_\odot$) star that could later explode as a core-collapse SN. It appears 
to be more consistent with a system dominated by a lower mass (below 8 M$_\odot$) giant star of mid-K spectral type.

\section{Bolometric light curves, temperatures, and radii} \label{Sect:bolom}

Using the light curves presented in Sect. \ref{Sect:lightcurves}, along with the distance and reddening values computed in Sect. \ref{Sect:hosts}, we can estimate the evolution of the luminosity 
through a blackbody fit to the spectral energy distribution (SED) at selected epochs. This procedure has been previously applied for other LRN studies \citep[e.g.][]{bla17,cai19,pas19b,pasto20}.

\begin{figure}
   \centering
   \includegraphics[width=9.2cm,angle=0]{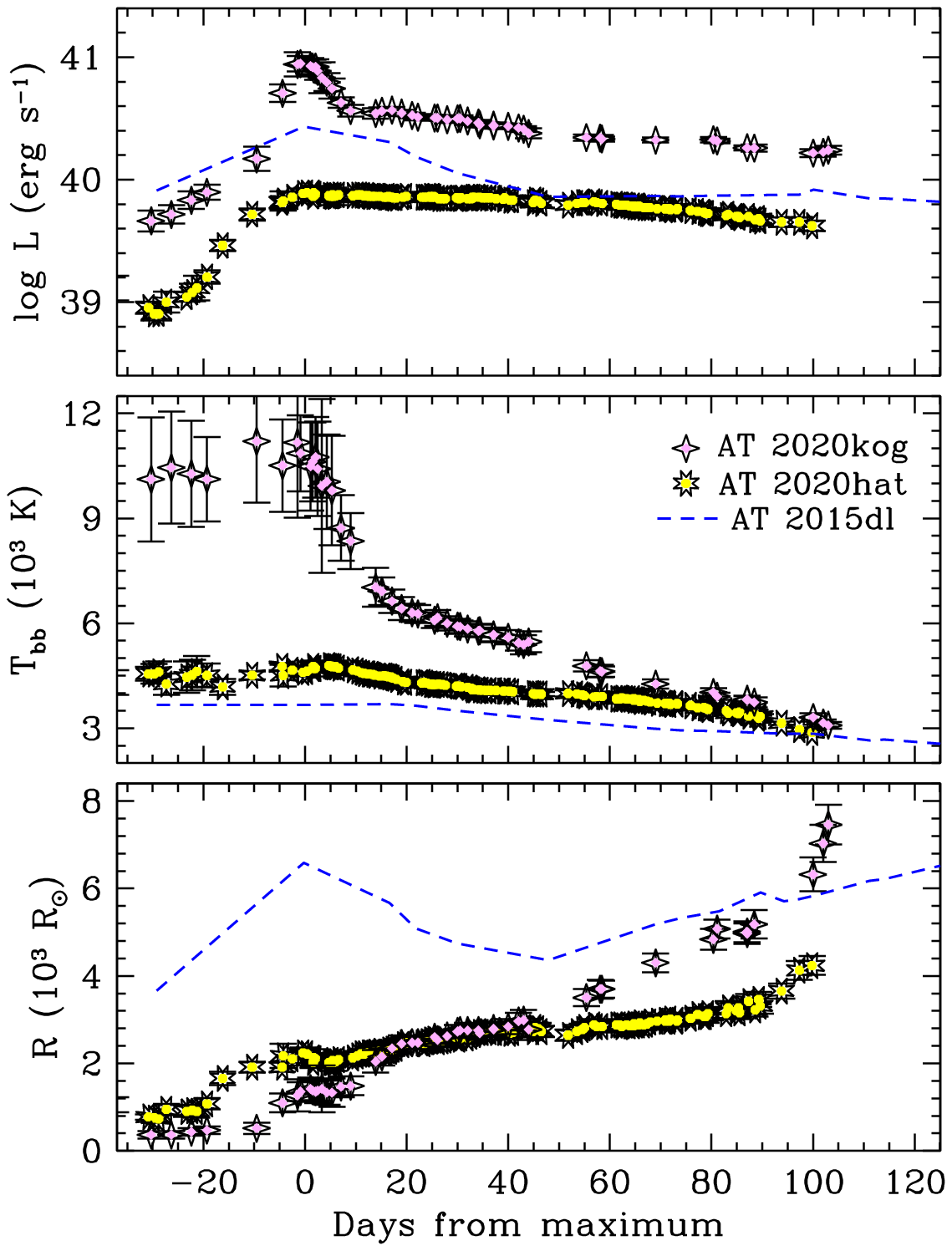}
   \caption{{\sl Top panel:} Bolometric light curves of AT~2020hat and AT~2020kog, compared with that of AT~2015dl/M101-2015OT1 \protect\citep{bla17}.
{\sl Middle panel:} Evolution of the blackbody temperature. {\sl Bottom panel:} Evolution of the radius at the photosphere.}
              \label{fig:bolo}
    \end{figure}

For AT~2020hat, the SED fits were computed by accounting for the contributions of the following bands: Sloan-$u$ (until phase $\sim$+50~d after maximum); Sloan-$g$ and ATLAS-$c$ (until phase 
$\sim$+100~d); Sloan-$r,i,z$ and ATLAS-$o$; and $J,H,K$ (from $\sim$+20 to $\sim$+100~d). For AT~2020kog, only the contribution of Sloan filter data have been considered, while the PanSTARRS-$w$ 
data were roughly approximated to Sloan-$r$. The flux contribution of the missing bands in the pre-outburst epochs were computed through the gross assumption that the source had the same colours 
as at maximum light. For both AT~2020hat and AT~2020kog, the SEDs are fairly well fitted by single blackbody functions, and the bolometric luminosity for each epoch was computed by integrating the 
inferred blackbody flux over the entire electromagnetic spectrum. The resulting bolometric light curves are compared in Fig. \ref{fig:bolo} (top panel) with that of AT~2015dl/M101-2015OT1 \citep{bla17}, 
an object whose luminosity is intermediate between those of the two LRNe discussed here.

The comparison with AT~2015dl is interesting because it has an intermediate luminosity between AT~2020hat and AT~2020kog. After the rise, AT~2020kog peaks at almost $9\times10^{40}$ erg s$^{-1}$, 
then rapidly declines until reaching the plateau at about $3\times10^{40}$ erg s$^{-1}$. Instead, AT~2020hat is about one order of magnitude fainter at maximum, peaking at  $8\times10^{39}$ erg s$^{-1}$, 
and then  its luminosity declines by almost a factor of 2 at phase $\sim$70~d after peak. AT~2015dl stays in the middle, although it has a much more prominent second maximum, while both AT~2020hat and 
AT~2020kog show a longer lasting plateau. An interpretation of the shapes of the LRN light curves is given in Sect. \ref{Sect:discussion}. 

The blackbody temperature ($T_{bb}$) evolution of the three LRNe is shown in Fig. \ref{fig:bolo} (middle panel). While AT~2020kog is initially hot (with temperatures $T_{bb}\sim10000-11000$ K), AT~2020hat 
is much cooler, having about 4700 K at maximum (hence comparable with the photosphere of an intermediate K-type star), with a temperature which is similar (slightly higher) than that of AT~2015dl. We note, 
however, that the pre-maximum temperatures of the two LRNe are highly uncertain due to the incomplete colour information regarding the pre-outburst epochs. With time, the temperature of AT~2020hat declines 
very slowly ($T_{bb}\sim3300$ K at about 3 months after maximum, which is consistent with that of an intermediate M-type star), while the temperature of AT~2020kog goes down much more rapidly, decreasing by 
a factor of two in two months and by a factor of three in the final epochs of the monitoring campaign.

Fig. \ref{fig:bolo} (bottom panel) shows the evolution of the photospheric radii of AT~2020hat and AT~2020kog, obtained through the Stefan-Boltzmann law. For both objects, the pre-outburst radius is relatively 
small at about 2 AU. Then, at the maximum light, the photospheric radius of AT~2020hat and AT~2020kog increases to 10.3 and 7 AU (hence, about 2200 and 1500  R$_\odot$), respectively. At $\sim$3 months after 
peak, the radii of AT~2020hat and AT~2020kog monotonically rise with time reaching about 16 AU (i.e. $\sim$3400 R$_\odot$) and 25 AU ($\sim$5400 R$_\odot$), respectively. At phases $>$+100 days, when the optical 
light curves decline very rapidly and the SED peak shifts towards the NIR domain, the photospheric radius is expected to rise significantly. In the case of AT~2020kog, the radius increases by 50 per cent in about 
two weeks. We also note that, at early phases, the photospheric radii of AT~2020hat and AT~2020kog are much smaller than that of AT~2015dl, but they become reasonably similar at late phases. In addition, while 
AT~2015dl shows a photospheric radius receding from 6500 to 4300~R$_\odot$ from the peak to $\sim$50 days after, this behaviour is not seen in AT~2020hat or AT~2020kog, which have a low-contrast early peak.

\section{Photometric  parameters of LRNe}  \label{Sect:discussion}

\begin{figure}
   \centering
   \includegraphics[width=9.4cm,angle=0]{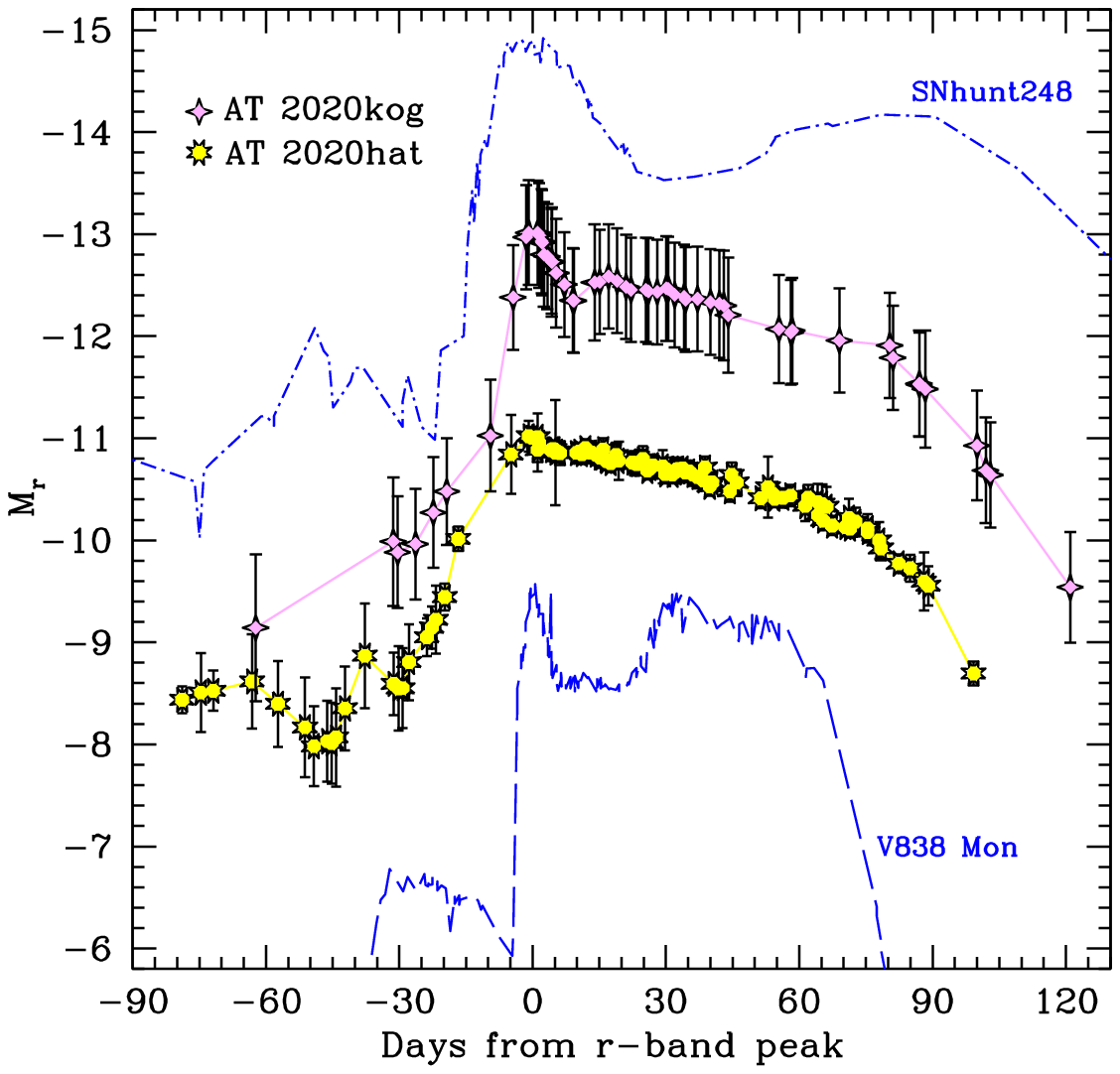}
   \caption{ Absolute light curves of AT~2020hat and AT~2020kog in the Sloan-$r$ bands, also showing the pre-outburst detections. As a comparison, we also show the $R$-band absolute 
light curves of the luminous SNhunt248 \protect\citep{kan15,mau15} and the relatively faint V838~Mon \protect\citep{mun02,gor02,kim02,cra03}, scaled to the AB mag system.}
              \label{fig:absolute}
    \end{figure}

AT~2020hat and AT~2020kog belong to the growing group of LRNe that were detected before the luminous outburst. The absolute magnitude of AT~2020hat in the last three months before the 
outburst ranges from $M_r \approx -8.0$ mag to $M_r \approx -8.9$ mag, with the light curve showing some fluctuations superposed on a general trend of luminosity rise (Fig. \ref{fig:absolute}). 
The pre-outburst phase of AT~2020kog was also less densely sampled because of its faintness. However, in the last $\sim$50~d before the outburst, the object monotonically brightens from 
$M_r \approx -9.1$ mag to $M_r \approx -10.4$ mag (see, Fig. \ref{fig:absolute}).

The pre-outurst light curves of AT~2020hat and, more marginally, AT~2020kog are similar to those observed in other LRNe, such as V1309~Sco \citep{tyl11} and M31-LRN2015 \citep[phases 1 to 3 as 
described in][see their figure 2]{bla20}. The initial slow luminosity rise observed in these LRNe accompanied by a low-contrast modulation in the light curve is usually interpreted as due to an 
increasing mass loss through the L2 point that generates an expanding photosphere \citep{pej17}. When this common envelope engulfs the binary, the luminosity decreases and the light curve shows 
a minimum, along with the disappearance of the superposed light-curve modulation. A further mass loss or shock interaction produced by the L2 outflow may cause a modest, later rebrightening, as 
those also observed in AT~2020hat and AT~2020kog. 

 This phase is followed by a major outburst, during which AT~2020hat and AT~2020kog reach $M_r \approx -11.0\pm0.1$ mag and $M_r \approx -13.2\pm0.5$ mag, respectively. This light curve peak is likely due 
to a hot, high-velocity gas outflow in polar direction following the system coalescence \citep{met17}. We note that AT~2020hat has a modest early peak, while it is much more prominent in AT~2020kog. The 
differences in the early LRN light curve shapes can be explained through the geometry of the system and the variety of masses involved in the process. This phase is then followed by a plateau, lasting 2-3 
months, which is reminiscent of that seen in type IIP SNe, rather than a broad, red second peak observed in the most luminous events \citep{pas19a}. Radiative shocks generated in the interaction between 
the newly expelled fast outflow and slower circumstellar material ejected before the coalescence power the plateau \citep{met17}. The material gathered through the shocks then produces a cool dense shell 
where dust may form in short time scales. The progressively increasing photospheric radius and the slow decline of temperature  observed in this phase (see Sect. \ref{Sect:bolom}) are fairly well explained 
with this scenario. Nonetheless, the mechanisms producing the heterogeneous photometric observables of LRNe are still under debate \citep[e.g.][]{mac17}.

Despite the uncertainties in the mechanism powering the light curves of LRNe, it is reasonable to assume that the ejected and total masses involved are key parameters to explain the widely heterogeneous 
photometric properties of these gap transients. \citet{pas19a} propose that individual features of the light curve (the $V$-band absolute magnitudes of the pre-outburst rise, the first peak, and the second 
peak or the plateau) are correlated with the intrinsic luminosity of the quiescent progenitor, while \citet{koc14} state that the light curve peak magnitudes are correlated with the total mass of the progenitor system.

   \begin{figure*}
   \centering
   \includegraphics[width=18.5cm,angle=0]{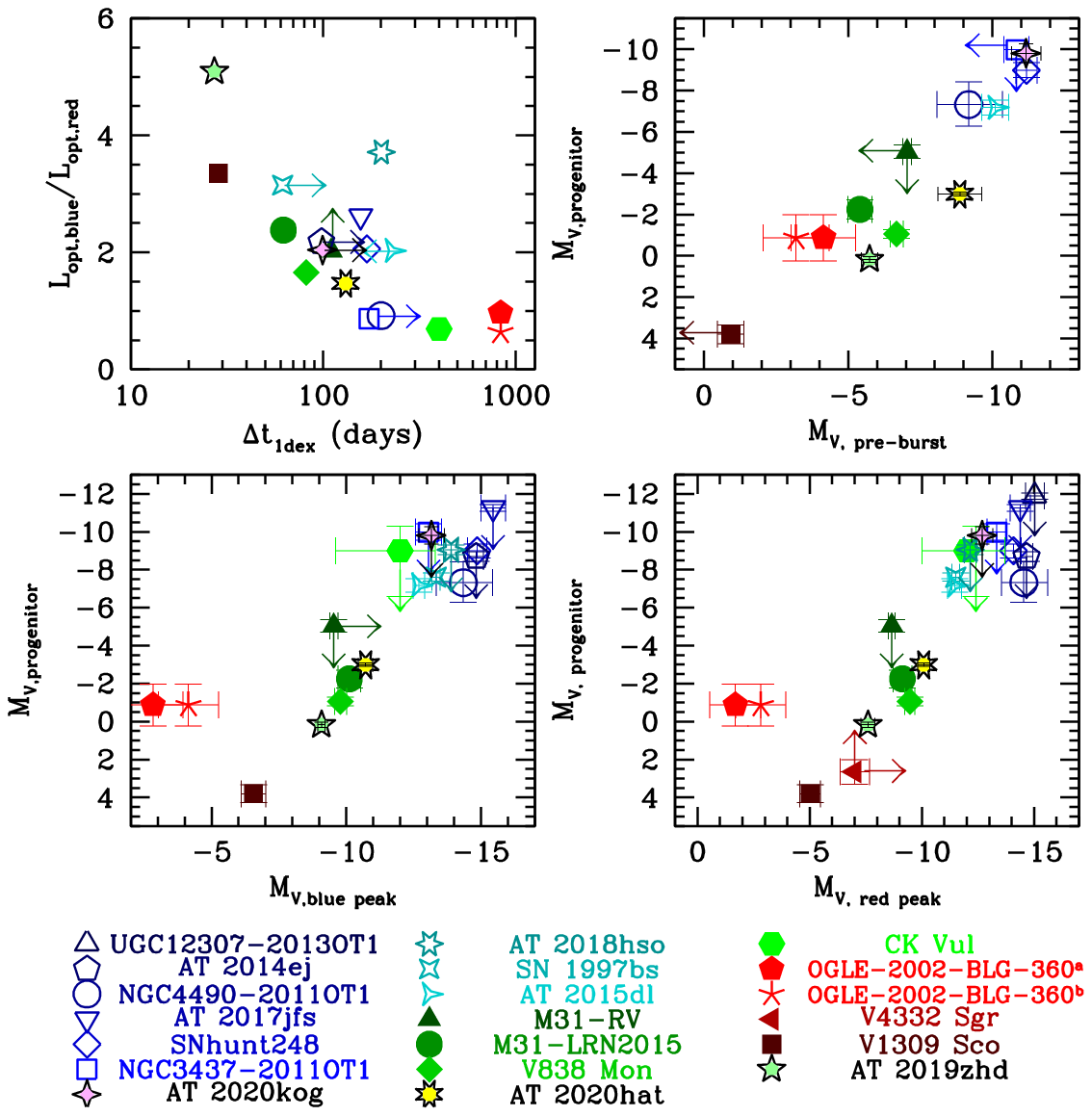}
   \caption{Correlations among photometric parameters in LRNe: ratio between blue and red  peak optical luminosities versus the period during which the LRNe have luminosities in the range 
of $0.1L_{peak}$ and $L_{peak}$ ({\sl Top-left panel}). The $V$-band absolute magnitude of the quiescent progenitor versus the  $V$-band absolute magnitude of the brightest point during the 
slow pre-outburst phase ({\sl top-right panel}), the $V$-band absolute magnitude of the first, blue peak ({\sl bottom-left panel}), and the $V$-band absolute magnitude of the second, red peak 
(or the plateau) ({\sl bottom-right panel}).}
              \label{fig:correlations}
    \end{figure*}

Unfortunately ultra-deep pre-explosion images of the host of AT~2020kog are not available, hence very few constraints can be placed on the luminosity of its progenitor. The stacked
Pan-STARRS images allowed us to only set detection limits for the progenitor of AT~2020kog down to $g=23.30$ and $r=22.97$ mag. The pre-outburst brightest Pan-STARRS $w$-band magnitude 
of AT~2020kog was converted to Johnson-$V$ through the relations given by \citet{ton12}, while Johnson-$V$ magnitudes of the two main peaks were obtained from the $g$-band magnitudes 
adopting the prescriptions of \citet{chr08} and the available colour information. In order to obtain the desired magnitudes of the pre-outburst maximum of AT~2020hat, along with those 
of the two main peaks, we used the available Sloan-$g$ and $r$ magnitudes, transformed to Johnson-$V$ following \citet{chr08}. When Sloan filter magnitudes were not available, they were 
estimated from the ATLAS-$c$ and $o$ magnitudes using  the relations of \citet{ton18} and \citet[][]{ton12} to account for the conversion from Pan-STARRS to Sloan photometry systems.
A similar approach was also adopted for other LRNe without a $V$-band observation at the crucial epochs discussed here \citep[see, also,][]{pas19a}. Following these prescriptions, from 
the stacked Pan-STARRS images, we infer an upper limit for the absolute magnitude of the quiescent progenitor of AT~2020kog of $M_V > -8.9$ mag.

As discussed in Sect. \ref{Sect:progenitor}, we can place much tighter constraints on the progenitor of AT~2020hat. Using the conversion formula between HST $F606W$ and Johnson-$V$ \citep{dol00}, 
the quiescent progenitor of AT~2020hat  would have a relatively faint absolute magnitude, that is $M_V=-2.99\pm0.09$ mag. For these two objects and the recent LRN AT~2019zhd \citep{pasto20}, we 
can compute  the absolute magnitudes of the brightest pre-outburst phase, and those of the blue peak and the red peak (or the plateau). Their photometric parameters can be compared with those of 
other LRNe from the literature in the updated version of the diagram originally presented in \citet{pas19a} (see Fig. \ref{fig:correlations}), with the inclusion of AT~2018hso \citep{cai19}, 
AT~2014ej from \citet{max20b}, AT~2019zhd \citep{pasto20}, and the two LRNe discussed in this paper. In the top-left panel, we also report where LRNe are located in the diagram showing the ratio 
between blue and red  peak optical luminosities versus the time during which the LRN light curves stay in the luminosity range between $0.1L_{peak}$ and $L_{peak}$. 

We note that the parameters of M31-LRN2015 \citep{wil15,kur15,lip17,bla20}, M31-RV \citep[][and references therein]{bos04}\footnote{The $V$-band light curve parameters of M31-RV have been fine-tuned 
including a few observations from \protect\citet{bry92}, which were not considered in \protect\citet{bos04}.}, and OGLE-2002-BLG-360 \citep{tyl13} have been updated with respect to those discussed in 
\citet[][]{pas19a}. In particular, as the light curve of OGLE-2002-BLG-360 is peculiar and shows a triple peak, we discuss two scenarios: Scenario (a), in which the first maximum corresponds to the 
brightest pre-merging detection, the second maximum is the blue peak, and the third one is the red peak; Scenario (b) assumes that the first light-curve maximum  is coincident with the blue peak, the 
second maximum is the red peak, while the third maximum is an unprecedented late optical re-brightening, which has never been observed in other LRNe thus far\footnote{To date, a late third light curve 
peak has only been observed in LRN AT~2017jfs, but in the NIR domain \citep{pas19b}.}. Finally, we have revised the distance estimate of the Galactic LRN CK~Vul, adopting the value reprted by 
\citet{ban20}\footnote{For CK~Per, \protect\citet{ban20} inferred a distance of $d=3.2^{1.3}_{-2.4}$ kpc and adopted a total line-of-sight absorption of $A_V = 2.47\pm0.45$ mag from \protect\citet{mar06}.}. 

While we confirm the existence of a correlation between the progenitor absolute magnitude and the light curve luminosity as discussed in \citet{pas19a}, in this paper, we also note another general 
trend among LRNe. In particular, objects with a larger blue peak to red peak luminosity ratio have light curves fading more rapidly in luminosity (Fig. \ref{fig:correlations}; top-left panel).

On the bright extreme of the LRN luminosity function (see Fig. \ref{fig:correlations}), we find extra-galactic events, such as NGC~4490-2011OT1 \citep{smi16,pas19a}, SNhunt248 \citep{kan15,mau15}, and 
AT~2015dl \citep{gor16,bla17,pas19a}. They are likely produced by massive, hot stellar systems, with masses of a few tens of M$_\odot$ \citep{smi16,bla17,mau18,pas19a}. The final outcomes of these 
high-luminosity LRNe are massive stars that will complete the nuclear burning cycles, finally exploding as core-collapse SNe. A significant fraction of core-collapse SNe are in fact expected to arise in 
close binary systems, including those producing a massive merger \citep[e.g.][]{zap17}. A massive stellar merger, for instance, was proposed to explain the peculiar circumstellar environment of SN~1987A 
and its blue supergiant progenitor \citep{mor07}. 

On the faint side of the LRN luminosity function, we have faint Galactic transients, such as V1309~Sco \citep{mas10,tyl11,wal12}, V4332~Sgr \citep{mar99,kim06,gor07}, and OGLE~2002-BLG-360 \citep{tyl13}. 
These transients are likely produced by cooler low mass systems, with a primary of about 1 M$_\odot$ and a secondary of 0.1M$_\odot$ (or even less). While a binary interaction may produce a single stellar 
merger \citep{tyl11}, the binary system producing an LRN may also survive after the common envelope ejection. After the expansion of the giant primary, the loss of orbital momentum reduces its separation 
from the low-mass companion. This leads to the ejection of the common envelope and the birth of a short-period binary, such as the WZ~Sge cataclysmic variables, which were formed by a white dwarf and a 
low-mass cool dwarf \citep{web93,kat15}. However, there are additional configurations that may produce LRN outbursts, including more complex multiple systems. For instance, V4332~Sgr was proposed to be a 
triple system, with the LRN ouburst being produced by a contact binary \citep{bar14}. Regardless of the evolutionary paths of their progenitor systems, the outcomes of lower-luminosity LRNe will likely 
end their life without producing an SN explosion.

In this framework, AT~2020kog can be regarded among the most luminous events generating massive mergers, while AT~2020hat has intermediate photometric properties similar to those of AT~2015dl. AT~2019zhd 
\citep[whose progenitor was over three mag fainter and redder than that of AT~2020hat,][]{pasto20} shares instead some similarity with other LRNe discovered in M~31, and more marginally with V838~Mon, hence 
the merger is likely a low to moderate-mass star, which is not expected to end its life as a core-collapse SN.

While light curve modelling \citep[e.g.][]{mac17,met17} has improved our knowledge on the LRN outburst mechanisms, very little is known about their progenitor systems. Strategies with high-cadence multiband 
observations are necessary to better understand the binary properties and the masses of the two components during the pre-merger phase of orbital instability, as has been done for V1309~Sco \citep{tyl11} 
and -- with less accuracy -- M31-LRN2015 \citep{bla20}. In addition, future deep images with high spatial resolution will verify our predictions as to the fate of the final merger, in particular through 
information provided by multi-domain observations, as imaging in a single domain \citep[e.g.][]{bon06,bon18} has often remained ambiguous.

\begin{acknowledgements}

We thank David Buckley for providing acquisition imaging taken with the Southern African Large Telescope (SALT), and Paolo Ochner for providing a few images taken with the
Copernico Telescope of Asiago (INAF-Osservatorio Astronomico di Padova). We are grateful to the referee V. Goranskij for carefully reading the manuscript and for the insightful comments.\\

MF gratefully acknowledges the support of a Royal Society – Science Foundation Ireland University Research Fellowship.\\
Research by SV and YD is supported by NSF grants AST–1813176  and AST-2008108.\\
Time domain research by DJS is supported by NSF grants AST-1813466, 1908972, and by the Heising-Simons Foundation under grant \#2020-1864.\\
YZC is supported in part by National Natural Science Foundation of China (NSFC grants 12033003, 11633002, 11325313, and 11761141001), National Program on Key Research and Development Project (grant no. 2016YFA0400803), and Ma Huateng Foundation.\\
DJ acknowledges support from the State Research Agency (AEI) of the Spanish Ministry of Science, Innovation and Universities (MCIU) and the European Regional Development Fund (FEDER) under grant AYA2017-83383-P.  DJ also acknowledges support under grant P/308614 financed by funds transferred from the Spanish Ministry of Science, Innovation and Universities, charged to the General State Budgets and with funds transferred from the General Budgets of the Autonomous Community of the Canary Islands by the Ministry of Economy, Industry, Trade and Knowledge.\\
SJS and KWS acknowledge funding from STFC Grants ST/P000312/1, ST/T000198/1 and ST/S006109/1.\\
EC and AF are partially supported by the PRIN-INAF 2017 with the project \textit{Towards the SKA and CTA era: discovery, localisation, and physics of transients sources} (P.I. M. Giroletti).\\
MS is supported by generous grants from  Villum FONDEN (13261,28021) and by a project grant (8021-00170B) from the Independent Research Fund Denmark.\\

This work is based on observations made with the Nordic Optical Telescope, operated by the Nordic Optical Telescope Scientific Association at the Observatorio del Roque de los Muchachos, La Palma, Spain, of the Instituto de Astrofisica de Canarias;
the Gran Telescopio Canarias (GTC), installed in the Spanish Observatorio 
del Roque de los Muchachos of the Instituto de Astrof\'isica de Canarias, in the Island of La Palma;  the Liverpool Telescope operated on the island of La Palma by Liverpool John Moores University at the 
Spanish Observatorio del Roque de los Muchachos of the Instituto de Astrof\'isica de Canarias with financial support from the UK Science and Technology Facilities Council; the Italian Telescopio Nazionale 
Galileo (TNG) operated on the island of La Palma by the Fundaci\'on Galileo Galilei of the INAF (Istituto Nazionale di Astrofisica) at the Spanish Observatorio del Roque de los Muchachos of the Instituto 
de Astrof\'isica de Canarias. This work makes also use of observations from the LCOGT network. \\

This work has made use of data from the Asteroid Terrestrial-impact Last Alert System (ATLAS) project. ATLAS is primarily funded to search for near earth asteroids through NASA grants NN12AR55G, 80NSSC18K0284, and 80NSSC18K1575; byproducts of the NEO search include images and catalogs from the survey area. The ATLAS science products have been made possible through the contributions of the University of Hawaii Institute for Astronomy, the Queen's University Belfast, the Space Telescope Science Institute, and the South African Astronomical Observatory, and The Millennium Institute of Astrophysics (MAS), Chile.\\

The Pan-STARRS1 Surveys (PS1) and the PS1 public science archive have been made possible through contributions by the Institute for Astronomy, the University of Hawaii, the Pan-STARRS Project Office, the Max-Planck Society and its participating institutes, the Max Planck Institute for Astronomy, Heidelberg and the Max Planck Institute for Extraterrestrial Physics, Garching, The Johns Hopkins University, Durham University, the University of Edinburgh, the Queen's University Belfast, the Harvard-Smithsonian Center for Astrophysics, the Las Cumbres Observatory Global Telescope Network Incorporated, the National Central University of Taiwan, the Space Telescope Science Institute, the National Aeronautics and Space Administration under Grant No. NNX08AR22G issued through the Planetary Science Division of the NASA Science Mission Directorate, the National Science Foundation Grant No. AST-1238877, the University of Maryland, Eotvos Lorand University (ELTE), the Los Alamos National Laboratory, and the Gordon and Betty Moore Foundation.\\

This publication is partially based on observations obtained with the Samuel Oschin 48-inch Telescope at the Palomar Observatory as part of the Zwicky Transient Facility project. ZTF is supported by the National Science Foundation under Grant No. AST-1440341 and a collaboration including Caltech, IPAC, the Weizmann Institute for Science, the Oskar Klein Center at Stockholm University, the University of Maryland, the University of Washington, Deutsches Elektronen-Synchrotron and Humboldt University, Los Alamos National Laboratories, the TANGO Consortium of Taiwan, the University of Wisconsin at Milwaukee, and Lawrence Berkeley National Laboratories. Operations are conducted by COO, IPAC, and UW.\\

This research has made use of the NASA/IPAC Extragalactic Database (NED) which is operated by the Jet Propulsion Laboratory, California Institute of Technology, under contract with the National Aeronautics and Space Administration. This publication made also use of data products from the Two Micron All Sky Survey, which is a joint project of 
the University of Massachusetts and the Infrared Processing and Analysis Center/California Institute of Technology, funded by NASA and the NSF. \\
This publication makes use of data products from the Two Micron All Sky Survey, which is a joint project of the University of Massachusetts and the Infrared Processing and Analysis Center/California Institute of Technology, funded by the National Aeronautics and Space Administration and the National Science Foundation.

\end{acknowledgements}

\end{document}